\documentclass[aps,prl,letterpaper,onecolumn,showpacs,superscriptaddress,floatfix,longbibliography]{revtex4}

\usepackage[english]{babel}
\usepackage{amsmath}
\usepackage{color}
\usepackage{epsfig}
\usepackage{graphicx}
\usepackage{bm}
\usepackage{times,amsmath,amssymb}
\usepackage{subfigure}
\usepackage{amsfonts}
\usepackage{amssymb}
\usepackage{amsmath}
\usepackage{color}
\usepackage{mathtools}
\usepackage{empheq}

\global\long\def\bra#1{\langle #1 |}
\global\long\def\ket#1{| #1 \rangle }

\global\long\def\braket#1#2{\left\langle #1|#2\right\rangle }

\global\long\def\abs#1{\left|#1\right|}

\global\long\def\al{\alpha}
\global\long\def\no{\nonumber}
\global\long\def\be{\beta}
\global\long\def\ga{\gamma}
\global\long\def\de{\delta}
\global\long\def\De{\Delta}

\global\long\def\th{\theta}

\global\long\def\en{E}

\global\long\def\si{\sigma}
\global\long\def\vfi{\gamma}

\global\long\def\bee{\begin{eqnarray}}
\global\long\def\eee{\end{eqnarray}}
\global\long\def\no{\nonumber}
\global\def\sh{{\sinh}}
\global\def\ch{{\cosh}}
\newcommand{\nc}{\newcommand}
\nc{\ir}{\mathrm{i}}
\nc{\dd}{\mathrm{d}}   
\nc{\eE}{\mathsf{e}}
\nc{\rP}{\mathrm{p}}
\nc{\rp}{p}
\def\multiplet{single particle multiplet}

\begin{document}

\title{Phantom Bethe roots in the integrable  open spin-$\frac12$ XXZ chain
}

\author{Xin  Zhang}
 \affiliation{Department of Physics,
  University of Wuppertal, Gaussstra\ss e 20, 42119 Wuppertal,
  Germany}
\author{  Andreas Kl\"umper} 
 \affiliation{Department of Physics,
  University of Wuppertal, Gaussstra\ss e 20, 42119 Wuppertal,
  Germany}
\author{Vladislav Popkov}
 \affiliation{Department of Physics,
  University of Wuppertal, Gaussstra\ss e 20, 42119 Wuppertal,
  Germany}

\begin{abstract}
We investigate special solutions to the Bethe Ansatz equations (BAE) for open
integrable XXZ Heisenberg spin chains containing phantom (infinite) Bethe
roots.  The phantom Bethe roots do not contribute to the energy of the Bethe
state, so the energy is determined exclusively by the remaining regular
excitations. We derive the phantom Bethe roots criterion and focus on BAE
solutions for mixtures of phantom roots and regular (finite) Bethe roots.  We
prove that in the presence of phantom Bethe roots, all eigenstates are split
between two invariant subspaces, spanned by chiral shock states. Bethe eigenstates
are described by two complementary sets of Bethe Ansatz equations for regular
roots, one for each invariant subspace.  The respective ``semi-phantom" Bethe
vectors are states of chiral nature, with chirality properties getting less
pronounced when more regular Bethe roots are added.  For the easy plane case
``semi-phantom" Bethe states carry nonzero magnetic current, and are
characterized by quasi-periodic modulation of the magnetization profile, the
most prominent example being the spin helix states (SHS). We illustrate our
results investigating ``semi-phantom" Bethe states generated by one regular
Bethe root (the other Bethe roots being phantom), with simple structure of the
invariant subspace, in all details.  We obtain the explicit expressions for
Bethe vectors, and calculate the simplest correlation functions, including the
spin-current for all the states in the \multiplet.

%For open system with FBR,  we prove that all eigenstates are split between two invariant subspaces, 
%spanned by chiral basises of shocks which contained either azimuthal  shocks or polar shocks, pertinent to easy-plane and %easy-axis regime respectively. 
%The states within the invariant subspaces generalize  spin helix states, featured recently in 
%the context of spin chains with strong boundary dissipation. 
\end{abstract}

%For open system with FBR,  we prove that all eigenstates are split between two invariant subspaces, 
%spanned by chiral basises of shocks which contained either azimuthal  shocks or polar shocks, pertinent to easy-plane and %easy-axis regime respectively. 
%The states within the invariant subspaces generalize  spin helix states, featured recently in 
%the context of spin chains with strong boundary dissipation. 

\maketitle

%\rev{PACs?}

\section{Introduction}

%\rev{first part  references to be added}

\bigskip

Contemporary experimental techniques allow to realize almost perfect
one-dimensional XXZ spin chains with adjustable anisotropy
\cite{2020NatureSpinHelix}.  The spin-$\frac{1}{2}$ XXZ chain, being an integrable
interacting many-body system, is one of the best studied paradigmatic models
in quantum statistical mechanics \cite{GaudinBook}.  Ongoing advances make the
XXZ model a source of inspiration and fascinating new discoveries, such as
finding a set of quasilocal conservation laws \cite{2013ProsenQuasilocal},
calculating finite temperature correlation functions
\cite{GoehmannKluemperSeel,BoosGoehmann}, and giving a major contribution to
the theory of finite-temperature quantum transport \cite{2020BertiniReview}.

In our letter \cite{PhantomShort} we have shown that SHS correspond to a novel
type of Bethe roots, phantom ``singular'' Bethe roots in the Bethe Ansatz
equations. These exotic Bethe roots describe excitations, which do not
contribute to the energy of the system.  We show that the presence of such
excitations leads to highly atypical chiral XXZ eigenstates carrying nonzero
spin currents and exhibiting periodic modulations of the magnetization density
profile.  We have established a criterion for the presence of phantom Bethe
roots, for both periodic and open boundaries, and investigated phantom Bethe
states formed exclusively from phantom Bethe excitations.

%In the present manuscript we focus on the situation when both phantom Bethe
%roots and regular Bethe roots (corresponding to usual thermal excitations) mix
%together. We treat the most challenging case of an open XXZ Hamiltonian with
%arbitrary non-diagonal boundary fields. Although the Bethe Ansatz equations for the spectrum  have been 
%constructed \cite{OffDiagonal},  little is known about the structure of the eigenstates.  
%We demonstrate that in presence of any number of
%phantom Bethe roots the Hilbert space of the system is split into two blocks,
%invariant with respect to the action of the Hamiltonian.  The ``splitting"
%theorem gives us a tool to study the structure of phantom Bethe states,
%belonging to each invariant subspace, and to show that ``semi-phantom" Bethe
%states retain the chiral character of ``fully-phantom" spin-helix states, as
%long as the number of regular Bethe roots involved remains small in comparison
%to the system size. In special cases the explicit form of phantom Bethe states
%and various observables including spin magnetization current can be calculated
%analytically.

In the present manuscript we treat the most challenging case of an open
  XXZ Hamiltonian with non-diagonal boundary fields. The Bethe
  Ansatz equations for the spectrum have been formally constructed
  \cite{OffDiagonal,Cao2013off,Zhang2015} for the general integrable open chain. The separation of
  variables method has been applied to this model in
  \cite{Niccoli2012,Faldella2014,Kitanine2014}.  And under a certain
  compatibility condition the so-called alternative modified Algebraic Bethe Ansatz
approach \cite{OffDiagonal03,Belliard} was successful. Despite these works
little is known about the solutions to the eigenvalue equations and even less is
  known about the structure of the eigenstates.  Here we demonstrate that in
  presence of any number of phantom Bethe roots the Hilbert space of the
  system is split into two blocks, invariant with respect to the action of the
  Hamiltonian. 
The condition for the splitting coincides with the condition needed for the applicability of the alternative modified Algebraic Bethe Ansatz \cite{OffDiagonal03,Belliard} and a conventional Baxter's $T$-$Q$ relation \cite{OffDiagonal}.

 %\rev{This situation arises under a condition of the type of
   % the compatibility condition necessary for the applicability of the
    %alternative modified Algebraic Bethe Ansatz.

We refer to our finding as the ``splitting theorem'' which gives us a tool to study the
  structure of phantom Bethe states, belonging to each invariant subspace, and
  to show that ``semi-phantom" Bethe states retain the chiral character of
  ``fully-phantom" spin-helix states, as long as the number of regular Bethe
  roots involved remains small in comparison to the system size. In special
  cases the explicit form of phantom Bethe states and various observables
  including spin magnetization current can be calculated analytically. 

The plan of the paper is the following.  After introducing the model we remind
of the concept of phantom Bethe roots, and derive the phantom Bethe roots
  existence criterion.  Next, we prove the theorem about the splitting of the
Hilbert space into two invariant chiral subspaces, and describe the basis states
spanning them.  In the final part of the manuscript we use the gained
knowledge to investigate the phantom Bethe states belonging to the invariant
subspace with dimension $N+1$ where $N$ is the length of the XXZ spin
chain. Details of the proofs are given in the Appendix.

\section{Phantom Bethe roots in  the open XXZ chain}

We consider the XXZ spin-$\frac{1}{2}$ chain with open boundaries.
The Hamiltonian reads \bee
H=\sum_{n=1}^{N-1}h_{n,n+1}+h_1+h_N,\label{Hamiltonian}
\eee
with 
\bee 
&&h_{n,n+1}=\sigma_n^x\sigma_{n+1}^x+\sigma_n^y\sigma_{n+1}^y+\ch\eta\sigma_n^z\sigma_{n+1}^z-\ch\eta\,I,\\[2pt]
&&h_1=\frac{\sh\eta}{\sh(\al_-)\ch(\be_-)}(\ch(\theta_-)\sigma_1^x+\ir\,\sh(\theta_-)\sigma_1^y+\ch(\al_-)\sh(\be_-)\sigma_1^z),\\[2pt]
&&h_N=\frac{\sh\eta}{\sh(\al_+)\ch(\be_+)}(\ch(\theta_+)\sigma_N^x+\ir\,\sh(\theta_+)\sigma_N^y-\ch(\al_+)\sh(\be_+)\sigma_N^z).
\eee
Here $\al_\pm$, $\be_{\pm}$ and $\th_{\pm}$ are boundary parameters. 
The system (\ref{Hamiltonian}) is integrable \cite{OffDiagonal} and its exact solutions are given by the off-diagonal Bethe Ansatz (ODBA) method \cite{OffDiagonal,Cao2013off,Zhang2015} and the separation of variables method \cite{Niccoli2012,Faldella2014,Kitanine2014}.  

Under generic open boundary conditions, the exact solutions of the system are given by an unconventional $T$-$Q$ relation with inhomogeneous term \cite{OffDiagonal,Cao2013off}, resulting in a set of %inhomogeneous%
BAE with $N$ so-called Bethe roots $\{\mu_1,\dots,\mu_N\}$
\begin{align}
&\frac{2 G(-\mu_j\!-\!\frac{\eta}{2})Q(-\mu_j\!-\!\frac{\eta}{2})}{\sinh( 2\mu_j\!+\!\eta )\sinh^{2N}(\mu_j\!+\!\frac{\eta}{2})}+\frac{2 G(\mu_j\!-\!\frac{\eta}{2})Q(\mu_j\!-\!\frac{\eta}{2})}{\sinh (2\mu_j\!-\!\eta)\sinh^{2N}(\mu_j\!-\!\frac{\eta}{2})}
=c\sinh( 2\mu_j),\qquad j=1,\ldots,N, \label{BAE}\\
&Q(u)=\prod_{k=1}^N\sinh(u-\mu_k-\frac{\eta}{2})\sinh(u+\mu_k-\frac{\eta}{2}),\nonumber\\
&G(u)=\prod_{\sigma=\pm}\sinh (u-\al_\sigma)\cosh (u-\beta_{\sigma})\,,\no\\
&c= \cosh\left[(N+1)\eta +\al_{-}+ \be_{-}+ \al_{+}+ \be_{+}\right]-\cosh(\th_{-}- \th_{+})\,. \label{Defc}
\end{align} 
All eigenvalues of the
Hamiltonian (\ref{Hamiltonian}) are classified by different sets of %$N$%
Bethe roots $\{\mu_j\}$ as
\begin{align}
&\en=\sum_{j=1}^N
\frac{4\sinh^2 \eta} {\cosh(2\mu_j)-\cosh\eta}+E_0,\label{E}\\[2pt]
&E_0 = -\sh\eta\left(\coth(\alpha_-)+\tanh(\beta_-)+\coth(\alpha_+)+\tanh(\beta_+)\right).\label{E0}
\end{align}

%where the roots satisfy the BAE of the form \cite{OffDiagonal,Zhang2015}

Note that unlike the periodic chain and the open chain with diagonal boundary fields which preserve the $ U(1)$ symmetry, here each eigenstate and the
corresponding eigenvalue are characterized by a set of Bethe roots $\{\mu_j\}$
with strictly $N$ members.  Typically, it is taken for granted that all
  $\{\mu_j\}_{j=1}^N$ are bounded, so that every Bethe root gives a nonzero
  contribution to the energy (\ref{E}). However, it was pointed out by us
  \cite{PhantomShort}, that unbounded ``phantom" solutions of BAE (\ref{BAE}) do
  exist, which lead to ``phantom" excitations not contributing to the
  energy. For completeness, below we give the definition and the derivation of
  the phantom Bethe roots existence criterion.  

\textit{Definition.} We shall call a Bethe root $\mu_p$ satisfying
(\ref{BAE}), a \textit{phantom} Bethe root, if it does not give a
contribution to the respective energy eigenvalue (\ref{E})
i.e.~if
\begin{align}
&{\rm Re}[\mu_p]=  \pm \infty.
\label{DefPhantomBR}
\end{align}

We assume that, out of $N$ Bethe roots, $N-M$ roots are phantom, 
\begin{align}
\mu_p=\infty + \ga_p, \ \ \ p=1,2,\ldots N-M,
\label{singBR}
\end{align}
where $\ga_p$ are some finite imaginary constants. The
more precise formulation of (\ref{singBR}) is $\mu_p=\mu_\infty + \ga_p$
with $\mu_\infty\to\infty$.
The remaining $M$ Bethe roots $\mu_{N-M+1}, \mu_{N-M+2},\ldots,\mu_{N} $ are
supposed to remain finite. In this situation, the BAE decouple for the phantom
roots and the regular roots.
Inserting (\ref{singBR}) into (\ref{BAE}), for
$1\leq p\leq N-M$ we obtain
\begin{align}
&\eE^W \prod_{k=1}^{N-M}2 \eE^{\ga_k-\ga_p} \sinh (\ga_p-\ga_k+\eta) +\eE^{-W} \prod_{k=1}^{N-M}2 \eE^{\ga_k-\ga_p}\sinh (\ga_p-\ga_k-\eta)
=2c,\label{singBAE}\\
&W=(M+1)\eta+\al_{-}+ \be_{-}+ \al_{+}+ \be_{+}.\no
\end{align}
Let us use the Ansatz
\begin{align}
&\ga_k=\ir \pi k/(N-M),  \label{AnsatzGak}
\end{align}
 and denote $\omega=\eE^{\ir \pi/(N-M)}$, so that $\omega^{N-M}=-1$ and
 $\eE^{\ga_k}=\omega^k$.  Then we can rewrite the first term on the LHS of (\ref{singBAE}) as
\begin{align}
&\prod_{k=1}^{N-M}\eE^{\ga_k-\ga_p} (2 \sinh (\ga_p-\ga_k+\eta)  )=\eE^{(N-M)\eta} \prod_{n=1}^{N-M}\left(1-\omega^{2n-2} \eE^{-2 \eta} \right)\no\\ 
&=\eE^{(N-M)\eta} \left(1 \!-\!\eE^{-2 \eta (N-M)} \right)\!=\! 2 \sinh ((N\!-\!M)\eta),\label{eq2}
\end{align}
where
we used the identity
\begin{align}
&\prod_{n=1}^{N-M}\left(1-\omega^{2n-2} z \right)=1-z^{N-M}\,,\label{polynomial}
\end{align}
as both sides are polynomials of degree $N-M$ in $z$, share the same zeros and
have identical $0$-th order coefficient. For $z=\eE^{-2 \eta}$ the RHS of (\ref{polynomial}) reduces to the term in brackets of line
(\ref{eq2}).  Analogously, we obtain
\begin{align}
&\prod_{k=1}^{N-M}2\eE^{\ga_k-\ga_j}  \sinh (\ga_j-\ga_k-\eta)=-2   \sinh((N-M)\eta).
\end{align}
The LHS of (\ref{singBAE})   can thus be rewritten as 
\begin{align}
&4 \sinh W \sinh ((N-M)\eta)= 2 \cosh((N-M)\eta+W ) - 2 \cosh((N-M)\eta -W )=2 c.
\label{singBAEfinal}
\end{align}
Recalling the definition of $W$ in (\ref{singBAE}) and  $c$ in (\ref{Defc})
we note that $2c=  2 \cosh( (N-M)\eta +W ) - 2 \cosh( \th_{-} - \th_{+} )$. In order 
to satisfy (\ref{singBAEfinal}) we must require  $\cosh( \th_{-} - \th_{+} )=
\cosh( (N-M)\eta -W )$, i.e. 
\begin{align}
\pm (\th_{+} - \th_{-}) &= (2M -N+1)\eta +\al_{-}+ \be_{-}+ \al_{+}+ \be_{+} \ \mod 2\pi \ir.
\label{ConditionPhantom}
\end{align}
Therefore, under condition (\ref{ConditionPhantom}), $N-M$ out of $N$ Bethe
roots in (\ref{BAE}) can be chosen phantom.  Integer $M$  naturally has the range
$0\leq M <N$.

To obtain the BAE for the $M$ remaining finite roots $x_j=\mu_{N-M+j}$, we
substitute (\ref{singBR}) into (\ref{BAE}) and take $j>N-M$.  The LHS of
(\ref{BAE}) contains factoring divergent terms, so that the finite
constant $2c$ on the RHS of (\ref{BAE}) can be neglected. The leading order
gives the final BAE \cite{OffDiagonal03,Rafael2003,Nepomechie2003}

\bee 
&&\left[\frac{\sh(x_j+\frac{\eta}{2})}{\sh (x_j-\frac{\eta}{2})}\right]^{2N}\prod_{\sigma=\pm}\frac{\sh(x_j-\alpha_\sigma-\frac{\eta}{2})}{\sh(x_j+\al_\sigma+\frac{\eta}{2})}\frac{\ch(x_j-\beta_\sigma-\frac{\eta}{2})}{\ch(x_j+\beta_\sigma+\frac{\eta}{2})}\no\\
&&=\prod_{k\neq j}^M\frac{\sh(x_j-x_k+\eta)\,\sh(x_j+x_k+\eta)}{\sh(x_j-x_k-\eta)\,\sh(x_j+x_k-\eta)},\quad j=1,\ldots,M.\label{BAE_1}
\eee
while the respective energy has contributions from the $M$ finite Bethe roots only,
\bee 
&&E=\sum_{j=1}^M\frac{4\sh^2\eta}{\ch(2x_j)-\ch\eta}+E_0.\label{energy_1}
\eee

 The condition (\ref{ConditionPhantom}) has been derived in
  \cite{OffDiagonal03,Nepomechie2003,Rafael2003,Cao2013off} as a restriction,
  under which a modified Algebraic Bethe Ansatz, based on special properties
  of Sklyanin's $K$-matrices, can be applied.  Alternatively, as is mentioned
  in \cite{OffDiagonal} the condition (\ref{ConditionPhantom}) gives a
  direct possibility to construct a conventional $T$-$Q$ relation without any
  inhomogeneous terms, see discussion around Eq.~(5.3.34) on pp.~145,148 in
  \cite{OffDiagonal}.  With both techniques (Algebraic Bethe Ansatz based on
  special properties of Sklyanin's $K$-matrices, and from the homogeneous
  $T$-$Q$ relation), the Bethe ansatz equations of the form (\ref{BAE_1}),
  (\ref{BAE_2}) for the spectrum can also be constructed.

The Hamiltonian $H$ is invariant upon the following substitutions
\begin{align}
\al_{\pm}&\rightarrow -\al_{\pm}\nonumber \\
\be_{\pm}&\rightarrow -\be_{\pm} \label{set2} \\
\th_{\pm}&\rightarrow  \ir \pi+ \th_{\pm}.\nonumber
\end{align}
Now Eq.~(\ref{ConditionPhantom}) will be mapped onto itself under substitutions (\ref{set2}) and $M \rightarrow \widetilde{M}=N-M-1$. Using substitutions (\ref{set2}) and letting $M\rightarrow \widetilde{M}$
in (\ref{BAE_1}) and (\ref{energy_1}), we obtain another set of BAE with $\widetilde M$
finite roots, namely
%and the respective eigenvalues.

\bee 
&&\left[\frac{\sh(x_j+\frac{\eta}{2})}{\sh (x_j-\frac{\eta}{2})}\right]^{2N}\prod_{\sigma=\pm}\frac{\sh(x_j+\alpha_\sigma-\frac{\eta}{2})}{\sh(x_j-\al_\sigma+\frac{\eta}{2})}\frac{\ch(x_j+\beta_\sigma-\frac{\eta}{2})}{\ch(x_j-\beta_\sigma+\frac{\eta}{2})}\no\\
&&=\prod_{k\neq j}^{\widetilde M}\frac{\sh(x_j-x_k+\eta)\,\sh(x_j+x_k+\eta)}{\sh(x_j-x_k-\eta)\,\sh(x_j+x_k-\eta)},\quad j=1,\ldots,{\widetilde M},\label{BAE_2}
\eee
while the respective energy has contributions from the ${\widetilde M}=N-M-1$ finite Bethe roots only,
\bee 
&&E=\sum_{j=1}^{\widetilde M}\frac{4\sh^2\eta}{\ch(2x_j)-\ch\eta}-E_0.\label{energy_2}
\eee

We remark that by our initial assumption about the existence of some phantom Bethe
  roots among the total of $N$ Bethe roots, the number $M$ of regular roots in
  (\ref{ConditionPhantom}) naturally takes the values $0\leq M< N$.  For
  condition (\ref{ConditionPhantom}) satisfied with $M=N$ it has been argued
  in \cite{Rafael2003} that the BAE set (\ref{BAE_1}) alone yields the full
  spectrum (of course with all Bethe roots regular).

For convenience, introduce the notation
$$\Delta=\ch\eta=\cos\gamma,\quad \eta=\ir\gamma.$$
When the constraint (\ref{ConditionPhantom}) holds, the hermiticity of the
Hamiltonian requires in the case $|\Delta|<1$ (the easy plane regime)
\bee {\rm Re[\al_{\pm}]=Re[\theta_{\pm}]= Re[\eta]=0},\,\, \rm Im[\be_{\pm}]=0\,\,{\mbox and }\,\,\be_+=-\be_-,\label{hermiticity_1}
\eee
and in the case $\Delta>1$ (the easy axis regime)
\bee {\rm Im[\al_{\pm}]=Im[\be_{\pm}]= Im[\eta]=0},\,\, \rm Re[\theta_{\pm}]=0\,\,{\mbox and }\,\,\theta_+=\theta_-\,\,mod \,\,2\ir\pi.\label{hermiticity_2}
\eee

Finally, for hermitian Hamiltonian (\ref{Hamiltonian}) the sign on the left
hand side of Eq.~(\ref{ConditionPhantom}) can be switched, by a
reparametrization $\al_\pm \rightarrow -\al_\pm$, $\eta \rightarrow -\eta$,
which leaves the Hamiltonian invariant.  Indeed, for $|\De|<1$, the left hand side of
Eq.~(\ref{ConditionPhantom}) switches sign under the reparametrization using $\be_{+}+ \be_{-}=0$, see (\ref{hermiticity_1}). For
$|\De|>1$ we have $\th_{+}= \th_{-}$ from (\ref{hermiticity_2}), so the
sign on the left hand side of Eq.~(\ref{ConditionPhantom}) is irrelevant.
Without loss of generality, we choose the ``$+$'' sign in
(\ref{ConditionPhantom}), yielding
\begin{align}
\th_{+} - \th_{-} &= (2M -N+1)\eta +\al_{-}+ \be_{-}
+ \al_{+}+ \be_{+} \quad  \mod 2\pi \ir.  \label{ConditionPhantomPlus} 
\end{align}

Below we formulate our main result, demonstrating a splitting of the Hilbert
space into two chiral invariant subspaces, if the condition
(\ref{ConditionPhantomPlus}) is fulfilled. 

%\rev{The theorem proved in the next section plays a key role in our study.}

\section{Mixture of phantom and regular roots: Splitting of the Hilbert space
  into two chiral invariant subspaces}
\label{sec::Splitting}

Here we show that under the Phantom Bethe roots (PBR) criterion 
(\ref{ConditionPhantomPlus}) the  Hilbert space splits into two  subspaces which are
invariant under the action of the open XXZ Hamiltonian (\ref{Hamiltonian})
and describe them. 

Define the following local vectors for each site $n$
\bee 
&&\phi_n(x)=\left(1,\,-\eE^{\th_-+\al_-+\be_-+(2x-n+1)\eta}\right),\label{local_stateBra}\\[2pt]
&&\tilde\phi_n(x)=\left(
\begin{array}{c}
	1\\[2pt]
	\eE^{-\th_--\al_--\be_-+(2x-n+1)\eta}
\end{array}
\right).\label{local_state}
\eee
Note the second component of these states depends on the position index $n$.
%The subscript $n$ in $\phi_n(x)$ and $\tilde{\phi}_n(x)$ indicates the quantum space.
Let us introduce  two families of factorized states parametrized by an integer
number $m$:
\begin{align}
&\Phi_{+}(n_1,\ldots, n_{m}) = \otimes_{k_1=1}^{n_1}\,\phi_{k_1}(M\!-\!m)\otimes_{k_2=n_1+1}^{n_2}\,\phi_{k_2}(M\!-\!m\!+\!1)\ldots \otimes_{k_{m+1}=n_m+1}^N\phi_{k_m}(M),\label{ResShockPlus} \\
%&f_k= F_1 +(k-1)\vfi\\
%&\be= \be_{-}; \quad F_1 = \pi  + i \al_{-}+ i \th_{-},\\
&1 \leq n_1 <n_2 \ldots <n_m\leq N,\qquad m=0,1,\ldots M,\no
\end{align}
and
\begin{align}
&\Phi_{-}(n_1,\ldots, n_{m}) = \otimes_{k_1=1}^{n_1}\,\tilde\phi_{k_1}(\widetilde M\!-\!m)\otimes_{k_2=n_1+1}^{n_2}\,\tilde\phi_{k_2}(\widetilde M\!-\!m\!+\!1)\ldots \otimes_{k_{m+1}=n_m+1}^N\tilde\phi_{k_m}(\widetilde{M}), \label{ResShockMinus} \\
%&f_k= F_1 +(k-1)\vfi\\
%&\be= \be_{-}; \quad F_1 = \pi  + i \al_{-}+ i \th_{-},\\
&1 \leq n_1 <n_2 \ldots <n_m\leq N,\qquad m=0,1,\ldots \widetilde{M},\no
\end{align}
where $ \widetilde{M}=N-M-1$.
\bigskip 

\textbf{Theorem.}
The open XXZ Hamiltonian, satisfying PBR criterion
(\ref{ConditionPhantomPlus}) with $0\leq M<N$ is block-diagonalized into two
complementary invariant subspaces $G_M^{+}$ and $G_{M}^{-}$ of dimensions
$\dim G_M^{+}= \sum_{k=0}^M \binom{N}{k}$ and $\dim G_M^-=2^N-\dim G_M^{+}$,
respectively.  $G_M^{+}$ is spanned by the family $\left\{ \Phi_{+}(n_1,\ldots, n_{m})\right\}_{m=0}^M$.  Subspace
$G_{M}^{-}$ is spanned by the family $\left\{
\Phi_{-}(n_1,\ldots, n_{m})\right\}_{m=0}^{ \widetilde{M}}$.  The eigenvalues of $H$
belonging to $G_M^{+}$ are given by the BAE  (\ref{BAE_1}), while those belonging to $G_M^{-}$ are given by (\ref{BAE_2}).

\bigskip
The proof of the ``invariance property'' of the subspaces $G_M^{\pm}$ for the
theorem is given in the Appendix.

For the rest of the theorem it remains to be demonstrated that the eigenstates of
$H$ belonging to the invariant subspaces $G_{M}^{+}$ and $G_{M}^{-}$ are
precisely those given by  BAE (\ref{BAE_1}) for $M$ and BAE (\ref{BAE_2}) for $ \widetilde{M}$,
respectively, leading further below to the sets of BAE (\ref{BAE;CBA}) and
(\ref{BAE;CBA2}).  Indeed, we observe precisely that  the BAE (\ref{BAE;CBA})
appear as consistency conditions when we construct the Bethe eigenstates via a
coordinate Bethe Ansatz for $M=1$ and $M=2$, and for larger $M>2$. The set of
basis states generated in the alternative modified Algebraic Bethe Ansatz
approach \cite{OffDiagonal03,Belliard} is equivalent to that given
by the theorem.

 Below, we demonstrate how this works for $M=1$, see section ``Mixtures of
 phantom and regular roots: ``semi-phantom" Bethe states".

\textit{Remark 1.}  The theorem is valid for an arbitrary Hamiltonian $H$ of
type (\ref{Hamiltonian}) satisfying (\ref{ConditionPhantomPlus}), whether it
is Hermitian or not. For the applications, we will consider hermitian $H$,
i.e.~with boundary parameters satisfying (\ref{hermiticity_1}) or (\ref{hermiticity_2}).

\textit{Remark 2.}  If one chooses $M$ outside of the range $[0,N-1]$ in  (\ref{ConditionPhantomPlus}) or (\ref{ConditionPhantom}), 
 a splitting of the Hilbert space will not occur. However, one can still argue that 
as a consequence, the whole spectrum will be governed by BAE of type  (\ref{BAE_1}) or  (\ref{BAE_2}) alone,
with the total number of regular Bethe roots  larger or equal to $N$, see elsewhere for details.

The proof of the theorem is our main result.

To evaluate observables in the phantom Bethe states we need further knowledge
about the Bethe amplitudes.  In the following we perform an exhaustive
analysis of phantom Bethe states for the $M=0$ and $M=1$ cases. Similar
results for $\widetilde M=0, 1$ hold after a substitution of the boundary parameters.

Below we shall explore the consequences of the theorem and construct phantom
Bethe states belonging to simple invariant subspaces, corresponding to a
mixtures of regular and phantom Bethe excitations.

\section{Spin helix states as ``perfect" phantom Bethe states }
By ``perfect" phantom Bethe states we mean the  Bethe states consisting
of exclusively infinite Bethe rapidities.  This case has been considered in detail
in \cite{PhantomShort}, and it corresponds to the choice $M=0$ in the phantom
Bethe roots existence criterion in (\ref{ConditionPhantomPlus}).

The invariant subspace $G_M^{+}$ for 
$M=0$ consists of a single state, the so-called spin helix state (SHS)
\bee
\langle SHS|=\phi_1(0) \phi_2(0)\cdots
\phi_N(0),\label{SHS}
\eee
which has the energy $E_0$ given by (\ref{E0}).
Another SHS corresponds to $G_M^-$ with $M=N-1$, or equivalently, ${\widetilde M}=0$,
\bee
 |SHS \rangle\!\rangle= \tilde\phi_{1}(0)  \tilde\phi_{2}(0)\cdots    \tilde\phi_{N}(0),\label{SHS-Gminus}
\eee
which has the energy $-E_0$. %Note that (\ref{SHS-Gminus}) is not a dual state to (\ref{SHS}), since they correspond to different Hamiltonians.
Despite being factorized states, SHS are rather nontrivial states of chiral nature,
characterized by periodic modulations of the polarization and large $O(1)$ magnetic current in the easy plane regime 
$\eta = \ir \ga$. The  magnetic currents for    (\ref{SHS}) and (\ref{SHS-Gminus}) are of opposite signs, reflecting the opposite chiralities,
\bee
j^z_{SHS} =\pm \frac{2\sin \ga}{\cosh ^2 (\be_{+})}, \label{jzSHS}
\eee
with $+$ and $-$ corresponding to  (\ref{SHS})  and (\ref{SHS-Gminus}) respectively.   
 Spin helix states can be prepared experimentally via coherent  \cite{2020NatureSpinHelix,2014HildSHS} and dissipative protocols \cite{2016PopkovPresilla,2017PopkovSchutzHelix}. 
Their chiral properties, and in particular their large $O(1)$ current (\ref{jzSHS})  
make them very different from typical eigenstates of many-body
interacting systems. This fact leads to singular features in the magnetization
current's dependence
on various system parameters in the proximity of ``phantom Bethe roots" manifolds  
in the dissipative protocols, see \cite{2020ZenoPRL,2020ZenoPRE}.

In the following we describe generalizations  of the spin helix states, and 
show that their distinct chiral features persist  also in presence of regular Bethe roots, as long 
as  phantom Bethe roots are present.

%and they play an important role in the  studies of
%quantum transport

Before proceeding to the description of Bethe states corresponding to
  mixtures of phantom and regular Bethe roots, we rewrite the BAE in a
  momentum representation which will be convenient for the further analysis.

\section{XXZ open spin-$\frac{1}{2}$ chain with phantom roots: momentum representation }

Following the traditional coordinate Bethe Ansatz method, we use the
  single particle quasi-momentum $\rP_j$ related to $x_j$ by
\bee 
\eE^{\ir{\rP}_j}=\frac{\sh\left(x_j+\frac{\eta}{2}\right)}{\sh\left(x_j-\frac{\eta}{2}\right)}\,.
\eee
The BAE (\ref{BAE_1}) become
\bee 
&&\eE^{2\ir N{\rP}_j}\prod_{\sigma=\pm}\frac{a_\sigma\,-\eE^{\ir{\rP}_j}}{1-a_\sigma\,\eE^{\ir{\rP}_j}} \,\,\frac{b_\sigma-\eE^{\ir{\rP}_j}}{1-b_\sigma\,\eE^{\ir{\rP}_j}}
=\prod_{\sigma=\pm}\prod_{k\neq j}^M\frac{1-2\ch\eta\,\eE^{\ir{\rP}_j}+\eE^{\ir{\rP}_j+\ir \sigma{\rP}_k}}{1-2\ch\eta\,\eE^{\ir \sigma{\rP}_k}+\eE^{\ir{\rP}_j+\ir \sigma{\rP}_k}},
\quad j\!=\!1,\ldots,M.\label{BAE;CBA}
\eee
where 
\bee 
a_{\pm}=\frac{\sh(\al_\pm+\eta)}{\sh(\al_{\pm})},\quad b_{\pm}=\frac{\ch(\be_\pm+\eta)}{\ch(\be_{\pm})},\label{def;ab}
\eee
and the selection rules of Bethe roots are
\bee
\eE^{\ir\rP_j}\neq \eE^{\pm\ir \rP_k}, \quad\eE^{\ir\rP_j}\neq \pm 1,  \quad \eE^{\ir{\rP_j}} \neq \eE^{\pm\eta}.
\label{SelectionRules}
\eee
The energy is given by
\bee 
&&E=4\sum_{j=1}^M\left(\cos({\rP}_j) -\ch\eta \right)+E_0.\label{Energy;CBA}
\eee

%Under the condition (\ref{ConditionPhantom}), we can also get another set of
%homogeneous BAE by replacing $\al_\pm$, $\be_\pm$ and $M$ in (\ref{BAE_1}) and
%(\ref{BAE;CBA}) with $-\al_\pm$, $-\be_\pm$ and $\widetilde M$ respectively,
%see \cite{PhantomShort}. 
The second set of BAE (\ref{BAE_2})
and the respective energy are rewritten in terms of single particle quasi-momenta as 
\bee
&&\eE^{2\ir N{\rP}_j}\prod_{\sigma=\pm}\frac{\tilde a_\sigma\,-\eE^{\ir{\rP}_j}}{1-\tilde a_\sigma\,\eE^{\ir{\rP}_j}} \,\,\frac{\tilde b_\sigma-\eE^{\ir{\rP}_j}}{1-\tilde b_\sigma\,\eE^{\ir{\rP}_j}}=\prod_{\sigma=\pm}\prod_{k\neq j}^{\widetilde M}\frac{1-2\ch\eta\,\eE^{\ir{\rP}_j}+\eE^{\ir{\rP}_j+\ir \sigma{\rP}_k}}{1-2\ch\eta\,\eE^{\ir \sigma{\rP}_k}+\eE^{\ir{\rP}_j+\ir \sigma{\rP}_k}},
\quad j=1,\ldots,\widetilde M,\label{BAE;CBA2}\\
&&\tilde a_{\pm}=\frac{\sh(\al_\pm-\eta)}{\sh(\al_{\pm})},\quad \tilde b_{\pm}=\frac{\ch(\be_\pm-\eta)}{\ch(\be_{\pm})},\label{def;ab2}\\
&&E=4\sum_{j=1}^{\widetilde M}\left(\cos({\rP}_j) -\ch\eta  \right)-E_0.\quad \label{Energy;CBA2}
\eee
We will show that solutions of (\ref{BAE;CBA}) and (\ref{BAE;CBA2}) constitute
the complete set of eigenstates and eigenvalues in the case with phantom Bethe
roots present, i.e.~under the criterion (\ref{ConditionPhantomPlus}).

\section{Mixtures of phantom and regular roots: ``semi-phantom" Bethe states}

Here we obtain generalizations of the spin-helix state (\ref{SHS}) for the case when all but one Bethe root are phantom, 
i.e.~there are
$N-1$ phantom Bethe roots and one regular Bethe root.  This situation arises at the manifold described by  
(\ref{ConditionPhantomPlus}) for $M=1$. 

We shall call the corresponding Bethe states semi-phantom Bethe states or, with  
 some abuse of notations,  as phantom Bethe states.
%, meaning that we consider cases where 
%the main contribution to the eigenstate is due to phantom Bethe roots \rev{???}. 

Here we construct explicit phantom Bethe eigenstates for $M=1$.

 The basis of the invariant subspace $G_1^{+}$, according to the theorem for the case $M=1$, is given by 
linearly independent vectors
 $\langle 0|,\langle 1|,\ldots ,\langle N|$ of the form
\bee
\langle n|=\eE^{n\eta}\phi_1(0)\cdots\phi_{n}(0)\phi_{n+1}(1)\cdots
\phi_N(1),\label{basis_1}
\eee
where the  prefactor $\eE^{n\eta}$ is introduced for  convenience. %and $x^+=x+2$.
The states  $\langle N|$ and  $\langle 0|$ are both SHS  of type  (\ref{SHS}),
with the same chirality, differing by an overall phase shift.
%$2 \eta/\ir$ \rev{$2\eta$ or $2\eta/\ir$?}
%\rev{(where the phase is the argument of the second component in  (\ref{local_stateBra})). }
The generic state  of the \multiplet\   $\langle n|$ is a state where the  pieces of  both 
SHS  are joined together at the link $n,n+1$ where  an additional  phase shift   occurs.
%the additional  phase shift $2 \eta/\ir$  occurs.

Any eigenstate of $H$ belonging to the invariant subspace  $G_1^{+}$  can be expanded as a linear combination of $\langle n|$ as 
\begin{align}
&\langle\Psi|=\sum_{n=0}^{N}\langle n|\,	f_n, \label{PsiM1}
\end{align} 
with the energy given by
\begin{align}
&E=4 \cos ({\rP_1}) - 4\cosh \eta+E_0.\label{EM1}
\end{align} 
where  ${\rP_1}$ satisfies the BAE (\ref{BAE;CBA}) and $E_0$ is given by (\ref{E0}). Obviously, (\ref{PsiM1})
predicts the existence of $N+1$ linearly independent Bethe vectors.

It is straightforward to verify that 
the action  of $\bra{n}H$ produces a linear combination of  
$\bra{n-1}$, $\bra{n}$, and $\bra{n+1}$ (see also the Appendix)
\begin{align}
&\bra{n} H = 2 \bra{n-1} + 2\bra{n+1} -d_{0} \bra{n},\no\\
&\bra{0} H = - \tilde{d}_0 \bra{0} + d_{-} \bra{1},\quad \bra{N} H = \tilde{d}_0 \bra{N} + d_{+} \bra{N-1}, \label{BoundaryM1}\\
&d_{0} = a_{+} +  b_{+} +    a_{-} +  b_{-}, \quad   \tilde{d}_0= a_{+} +  b_{+} -    a_{-} -  b_{-},\no\\
& d_\pm = 2 -2  a_{\pm}\,  b_{\pm}. \label{dplusminus}
\end{align}
Inserting the above equations into the eigenvalue problem $ \langle\Psi| H =E \  \langle\Psi|$
gives the following identities
\bee
&&f_1=\left(2\cos(\rP)-a_--b_-\right)f_0,\label{M1;b1}\\[2pt]
&&%\frac{a_-\,b_--1}{a_-\,b_-}\,
(1-a_-\,b_-)f_0+f_2=2\,\cos(\rP)\,f_1,\label{M1;b2}\\[2pt]
&&f_{n-1}+f_{n+1}=2\,\cos(\rP) \,f_n,\quad 2\leq n\leq N\!-\!2,\label{M1;b3}\\[2pt]
&&%\frac{a_+\,b_+-1}{a_+\,b_+}\,
(1-a_+\,b_+)f_{N}+f_{N-2}=2\,\cos(\rP)\,f_{N-1},\label{M1;b4}\\[2pt]
&&f_{N-1}=\left(2\cos(\rP)-a_+-b_+\right)f_{N}.\label{M1;b5}
\eee
where 
$a_\pm, b_\pm$ are given by (\ref{def;ab}), and ${\rP}\equiv {\rP_1}$. 

We use the Ansatz
\bee 
\begin{aligned}\label{M1;Ansatz}
&f_{n}=A_+\eE^{\ir n\rP}+A_-\eE^{-\ir n\rP},\quad 1\leq n\leq N-1,\\
&f_0=\frac{1}{1-a_-b_-}\left(A_++A_-\right),\\
%\frac{a_-\,b_-}{a_-\,b_--1}
&f_N=\frac{1}{1-a_+b_+}\left(A_+\eE^{\ir N\rP}+A_-\eE^{-\ir N\rP}\right),
\end{aligned}
%\frac{a_+\,b_+}{a_+\,b_+-1}
\eee
which satisfies Eqs.~(\ref{M1;b2}) - (\ref{M1;b4}). To satisfy (\ref{M1;b1}) and (\ref{M1;b5}), we need 
\bee 
&&\frac{A_+}{A_-}=-\frac{a_--\eE^{\ir\rP}}{1-a_-\,\eE^{\ir\rP}}\,\frac{b_--\eE^{\ir\rP}}{1-b_-\,\eE^{\ir\rP}},\label{M1;A+-}\\
&&\frac{A_-}{A_+}= -\,\eE^{2\ir N\rP}\,\frac{a_+-\eE^{\ir\rP}}{1-a_+\,\eE^{\ir\rP}}\,\frac{b_+-\eE^{\ir\rP}}{1-b_+\,\eE^{\ir\rP}}.\label{M1;A-+}
\eee
Multiplying the above equations, we retrieve BAE (\ref{BAE;CBA}) for $M=1$
\begin{align}
\eE^{2\ir N\rP}\,\frac{a_+-\eE^{\ir\rP}}{1-a_+\,\eE^{\ir\rP}}\,\frac{b_+-\eE^{\ir\rP}}{1-b_+\,\eE^{\ir\rP}}\,\frac{a_--\eE^{\ir\rP}}{1-a_-\,\eE^{\ir\rP}}\,\frac{b_--\eE^{\ir\rP}}{1-b_-\,\eE^{\ir\rP}}=1, \label{BAE-M1}
\end{align}
which thus serves as a 
compatibility condition of Eqs.~(\ref{M1;A+-}), (\ref{M1;A-+}). 
To sum up, the Bethe vectors of the \multiplet\  (invariant subspace $G_1^{+}$) are 
given by (\ref{PsiM1}) with $f_0,\ldots,f_{N}$ being simplified to
\bee 
&&f_n=A_+\eE^{\ir n\rP}+A_-\eE^{-\ir n\rP}=\sin(n\rP+\al),\qquad n=1,\ldots,N-1,\no\\
&&f_0 = \frac{1}{1-a_-b_-} \sin \al, \quad f_N = \frac{1}{1-a_+b_+} \sin (N \rP +\al), \label{f;alpha}\\ 
&&\al=\frac{1}{2\ir}\ln\left(\frac{a_--\eE^{\ir\rP}}{1-a_-\,\eE^{\ir\rP}}\,\frac{b_--\eE^{\ir\rP}}{1-b_-\,\eE^{\ir\rP}}\right).\label{alpha}
\eee
and ${\rP}$ satisfying BAE (\ref{BAE-M1}).

\textbf{Special cases}: {For the generic case, the structure of the invariant
  subspace $G_1^{+}$ spanned by $\{\bra{n}\}_{n=0}^N$ is a fully connected one; meaning
  that with a repeated action $\bra{n}H$, $\bra{n}H^2$, etc.~on any basis
  vector $\bra{n}$ one gets the full basis.  Consequently, any eigenvector
  from the \multiplet\ has nonzero components of all basis vectors, and all
  phantom Bethe eigenvectors are given by Eqs. (\ref{M1;Ansatz}) -
  (\ref{BAE-M1}). This remains true also for higher invariant subspaces
  $G_M^{\pm}$ which are generically fully connected.  However, in special
  cases, the invariant subspaces $G_M^{\pm}$
  may have additional internal structure, i.e.~contain invariant subspaces of
  smaller sizes.  This case of further partitioning, can already be
  illustrated on our example of $G_1^{+}$, which has internal structure if one
  or both coefficients $d_\pm$ in (\ref{dplusminus}) vanish, as is illustrated
  below.  }

(\textbf{\romannumeral1}) $d_{-}=0$. This happens for { $a_{-}b_{-}=1$, for which
 a one-dimensional invariant subspace of
  $G_1^{+}$ containing just one state, the basis state $\bra{0}$,
  appears, see (\ref{BoundaryM1}). Consequently, $\bra{0}$ itself is a Bethe eigenvector
  with eigenvalue $ -\tilde{ d}_0$, while the remaining $N$ phantom Bethe
  eigenvectors have nonzero components of all basis vectors, given by the BAE
  which now has $N$ solutions instead of $N+1$ solutions and $\al=0$.  }

(\textbf{\romannumeral2}) {$d_{+}=0$.} This happens for $a_+ b_+=1$ with similar
structures as in the $d_{-}=0$ case.  The SHS $\langle N|$ is an eigenstate of
$H$ with corresponding eigenvalue $\tilde d_0$. The other $N$
eigenstates are given by Eqs.~(\ref{M1;Ansatz}) , (\ref{BAE-M1}) and
(\ref{f;alpha}) with $\al=-N\rP$.

(\textbf{\romannumeral3}) {$d_{+}=d_{-}=0$. When $a_\pm b_\pm=1$,
  i.e.~$\ch(\al_\pm+\be_\pm+\eta)=0$ (see Eq.~(\ref{Iden-2})), $G_1^{+}$
  contains two one-dimensional invariant subspaces, $\bra{0}$ and $\bra{N}$,
  as follows from (\ref{BoundaryM1}). Consequently $\bra{N}$ and $\bra{0}$ become
  the eigenvectors of $H$ with eigenvalues $ \tilde{ d}_0$ and $-\tilde{ d}_0$,
  respectively. The remaining $N-1$ phantom Bethe states are given by BAE
  (\ref{BAE-M1}) which acquire the remarkably simple form
\begin{align}
\eE^{2\ir N\rP}=1\,, \label{BAE-iii}
\end{align}
yielding  $N-1$ real solutions
\bee 
\rP=\frac{m\pi}{N},\quad m=1,\ldots,N-1\,,
\eee
where due to  (\ref{SelectionRules}), $\rP=0,\pi$ are not allowed.
 }
From Eqs.~(\ref{M1;A+-}) - (\ref{alpha}), we find
\bee 
&&\al=0,\qquad f_n=\sin(n\rP),\qquad n=1,\ldots ,N-1,\no\\
&&f_0=\lim_{a_-b_-\to 1}\frac{\sin(\al)}{1-a_-\,b_-}=\frac{\sin(\rP)}{2\cos(\rP)-a_--b_-},\no\\
%\lim_{a_-b_-\to 1}\frac{\eE^{-\ir\al_0}}{2\ir(1-a_-\,b_-)}(\eE^{2\ir\al_0}-1)=\frac{\eE^{2\ir\rP}-1}{2\ir(1-a_-\eE^{\ir\rP})(1-b_-\eE^{\ir\rP})},\no\\
&&f_N=\lim_{a_+b_+\to 1}\frac{\sin(N\rP\!+\!\al)}{1-a_+\,b_+}=\frac{\sin((N\!-\!1)\rP)}{2\cos(\rP)-a_+-b_+}.
%\lim_{a_+b_+\to 1}\frac{\eE^{-\ir N\rP-\ir\al_0}}{2\ir(1-a_+\,b_+)}(\eE^{2\ir N\rP+2\ir\al_0}-1)=\frac{\eE^{\ir N\rP}(1-\eE^{2\ir\rP})}{2\ir(a_+-\eE^{\ir\rP})(b_+-\eE^{\ir\rP})}.
\eee

%The other two solutions are $\rP=-\ir \ln(a_\pm)$. When $\rP=-\ir \ln(a_-)$, $f_0\to 0$ and the eigenstate $\langle\Psi|$ becomes the SHS $\langle 0|$.  The solution $\rP=-\ir \ln(a_+)$ gives another SHS $\langle N|$.

\section{Properties of phantom Bethe vectors  for $M=1$.}

\textbf{Distribution of regular Bethe roots in the \multiplet.}  As expected,
there are $N+1$ physical solutions of the BAE (\ref{BAE;CBA}) for one
quasi-momentum ${\rP}\equiv\rP_1$ ($M=1$). These solutions for $\rP$ and $\al$ are
  denoted by $p_j$ and $\al_j$ with $j=1,..., N+1$. We refer to the sets
  $\{{p_j}\}_{j=1}^{N+1}$ and $\{{\al_j}\}_{j=1}^{N+1}$ of solutions as the \multiplet.

Note that for a hermitian Hamiltonian $H$ it
follows from (\ref{EM1}) that ${\rP}$ can be either real or purely imaginary. In the following we shall concentrate on the easy plane case
$|\De|<1$ which is physically more interesting since it produces eigenstates
with multiple windings of the magnetization vector along the chain for large
systems, see Fig.~\ref{Fig-profiles}.

It can be shown (see Appendix) that in the \multiplet\  $\{{\rp_j}\}_{j=1}^{N+1}$
there are $0,1$ or $2$ purely imaginary ${\rp_j}$ solutions depending on the system size
$N$ and on the values $a_\pm,b_\pm$; the remaining $N+1$, $N$ or $N-1$ solutions are real.

\begin{figure}[tbp]
\centerline{
\includegraphics[width=0.48\textwidth]{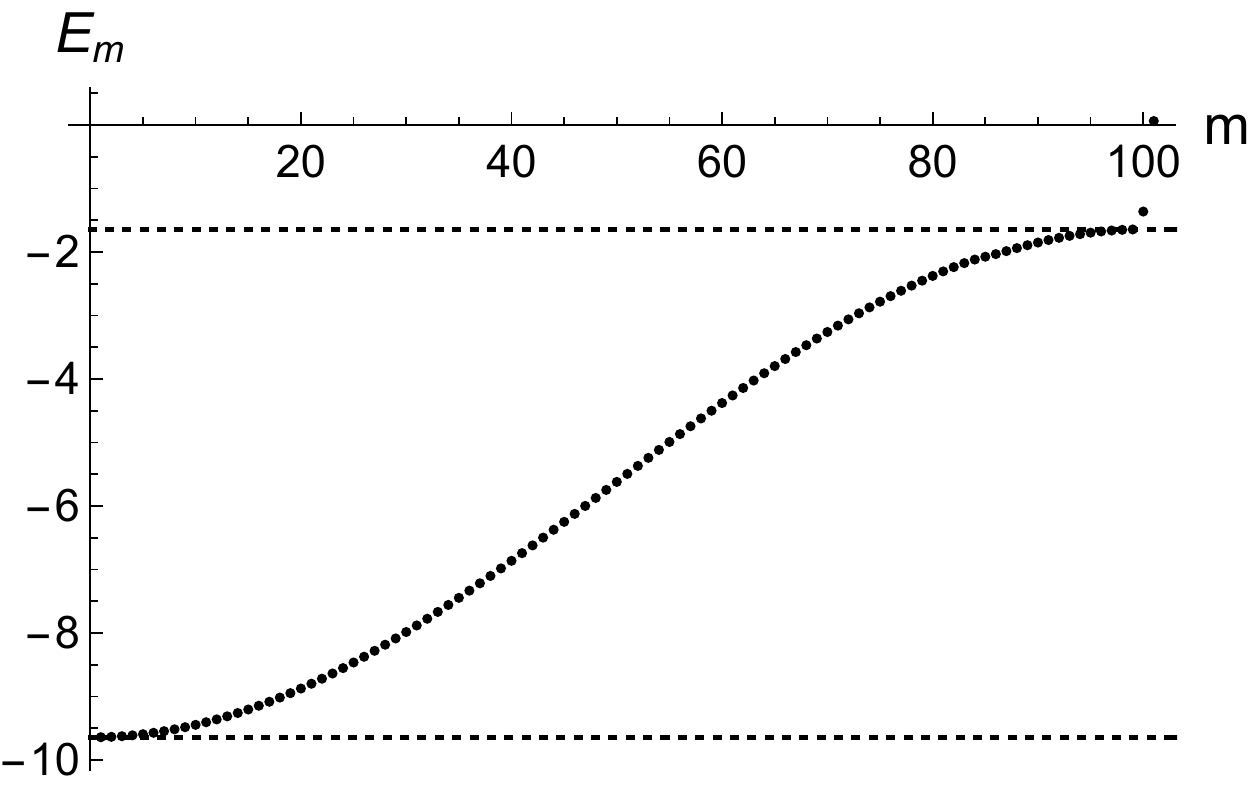}
}
\caption{ Energy levels $E_m=E_0 - 4 \De + 4 \cos ({p_m})$ plotted versus $m$
  of a \multiplet\ for $M=1$.  Dotted lines show strict lower and upper bounds
  $E_0- 4 \De -4$ and $E_0- 4 \De +4$, $E_0$ given by (\ref{E0}), for energies
  $E_m$ of real ${\rp_m}$ solutions. Two separate points representing the
  largest energies correspond to two imaginary ${\rp_m}$ solutions.
  Parameters are: $N=100, \ \De = \cos (\pi/6.2) \approx 0.874$, $\al_{+}= 0.3
  \ir$ , $\al_{-}= 0.7 \ir$, $\be_{+}= - \be_{-}=1.5$. The structure of selected
  phantom Bethe states belonging to the \multiplet\ is shown in Fig.~\ref{Fig-knots}.
The values of  $\theta_\pm$ are arbitrary 
imaginary parameters satisfying  (\ref{ConditionPhantom}).}
\label{Fig-energies}
\end{figure}

It can be argued that in the thermodynamic limit $N \gg 1$ the points $\tau_j
= \eE^{\ir\rp_j}$ for real ${\rp_j}$ densely populate the upper unit semicircle. Let us order
the real ${p_m}$ in order of increasing energy $E_m$. %The difference between the nearest quasi-momentum being ${\rp_{j+1}}-{\rp_{j}} \approx \pi/N$. In addition,
One finds $\min_j {\rp_j} = |O(1/N)|$
and $\max_j {\rp_j} =\pi - |O(1/N)|$. Thus, the energies of the real members
of the \multiplet\ $\{{\rp_j}\}_{j=1}^{N+1}$ densely populate the interval of
energies $E_j \in \left(E_0\!-\!4\Delta\!-\!4, E_0 \!-\!4\Delta\!+\!4\right)$, see Fig.~\ref{Fig-energies}. 

\begin{figure}[tbp]
\centerline{
\includegraphics[width=0.5\textwidth]{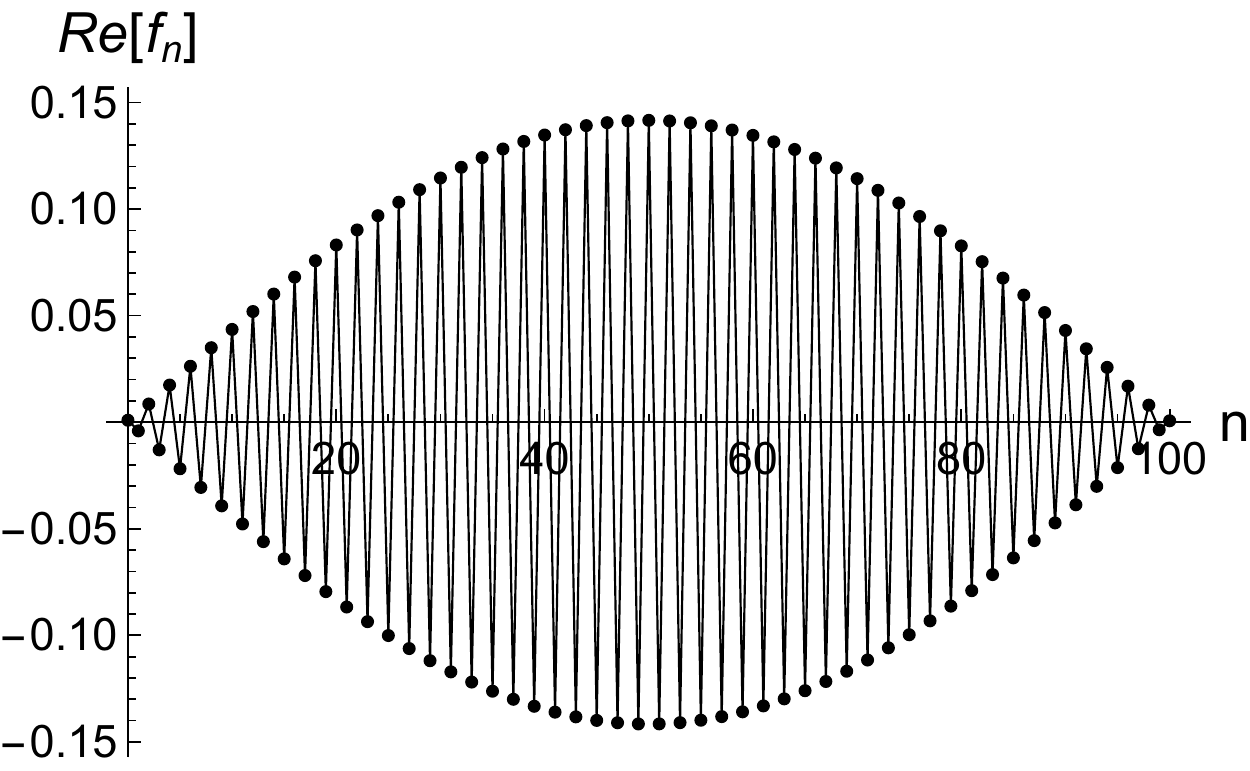}
\includegraphics[width=0.5\textwidth]{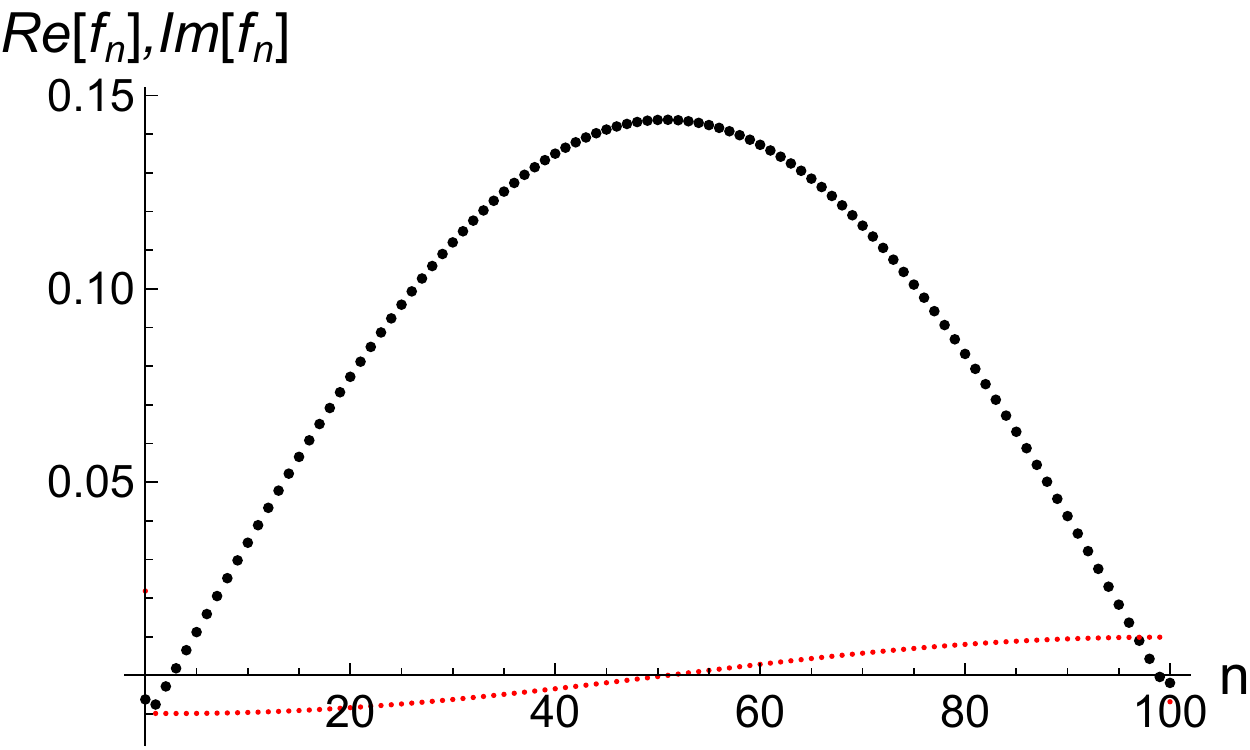}
}
\centerline{
\includegraphics[width=0.5\textwidth]{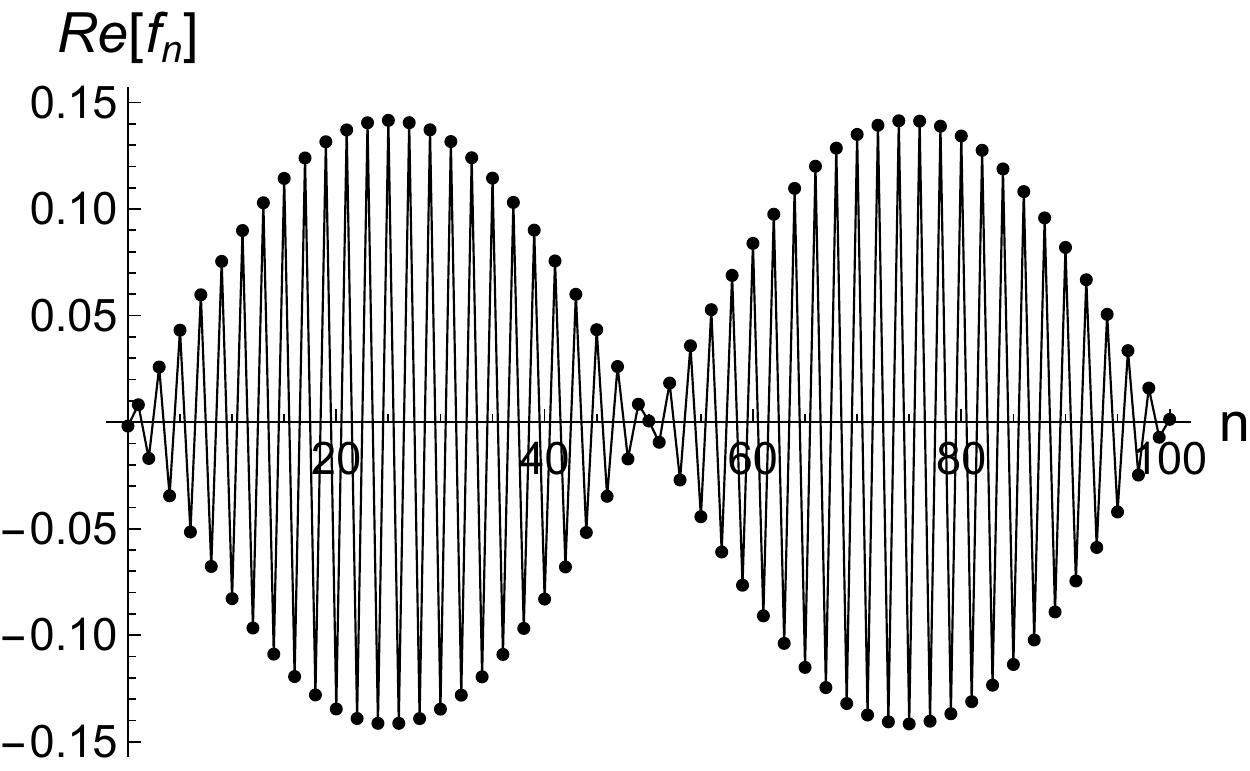}
\includegraphics[width=0.5\textwidth]{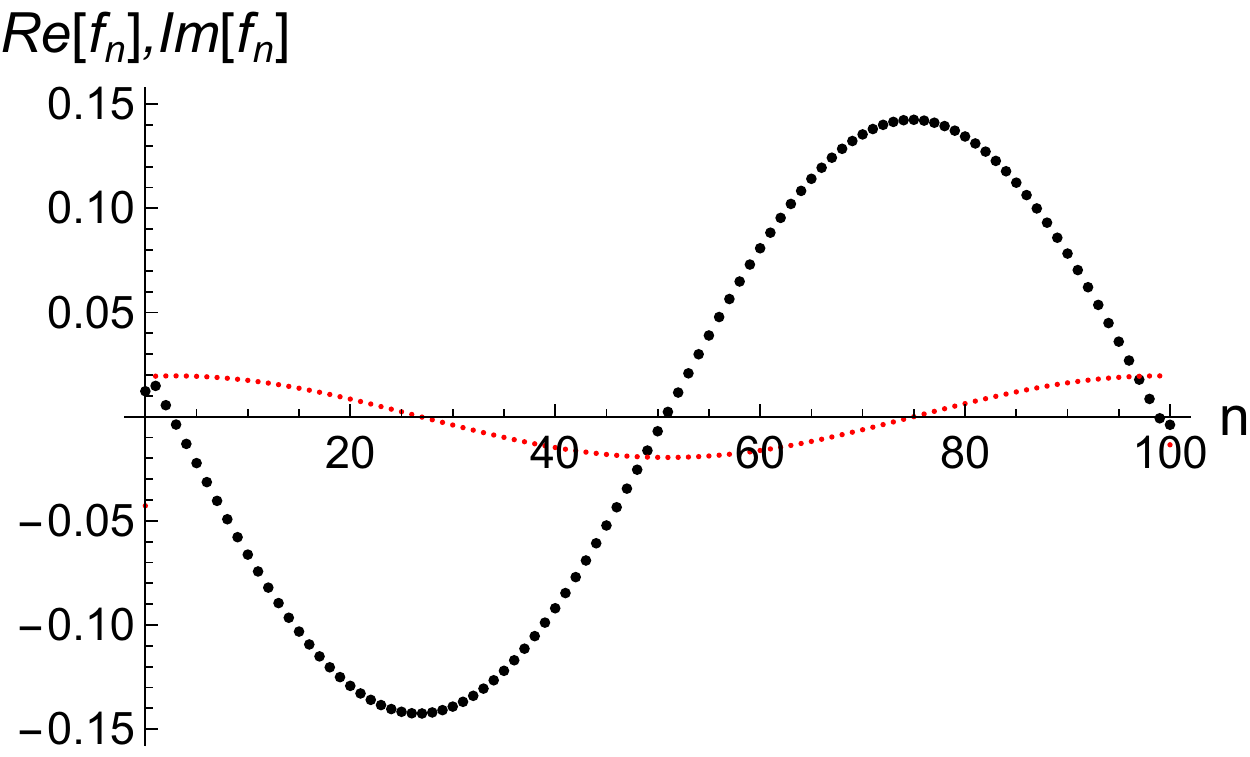}
}
\centerline{
\includegraphics[width=0.5\textwidth]{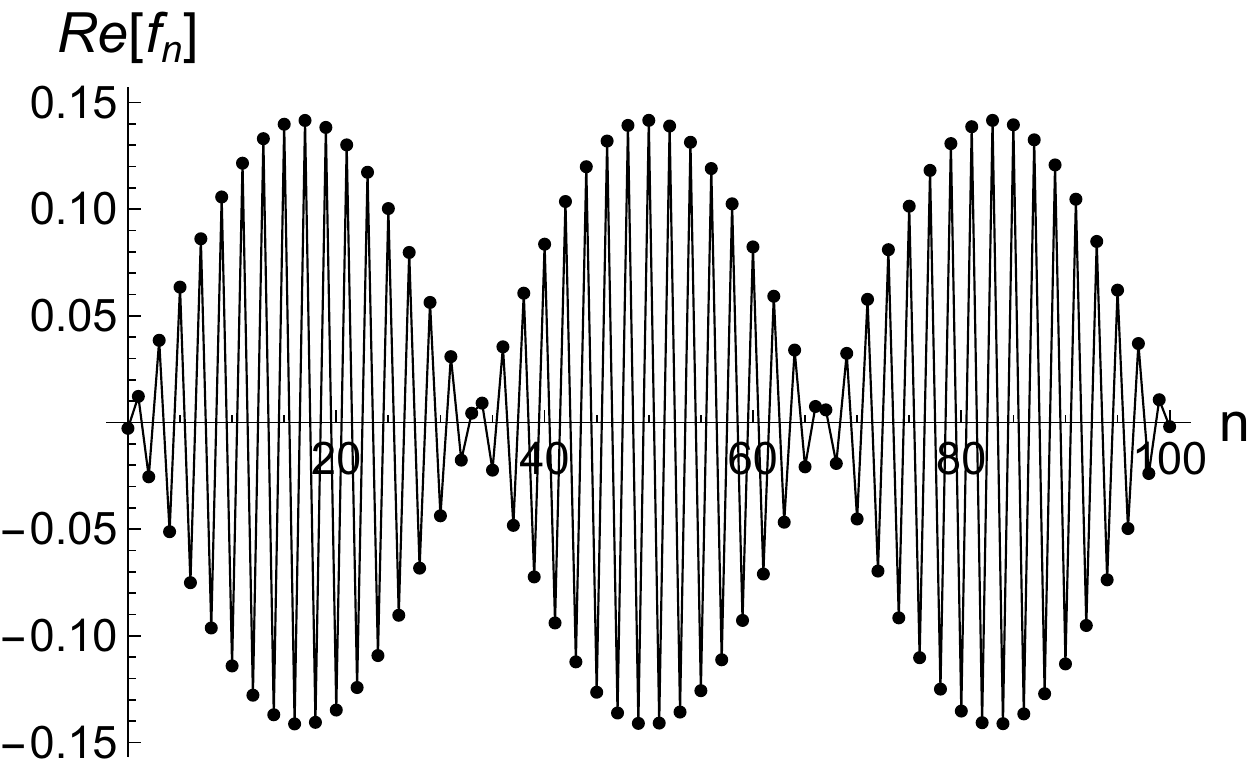}
\includegraphics[width=0.5\textwidth]{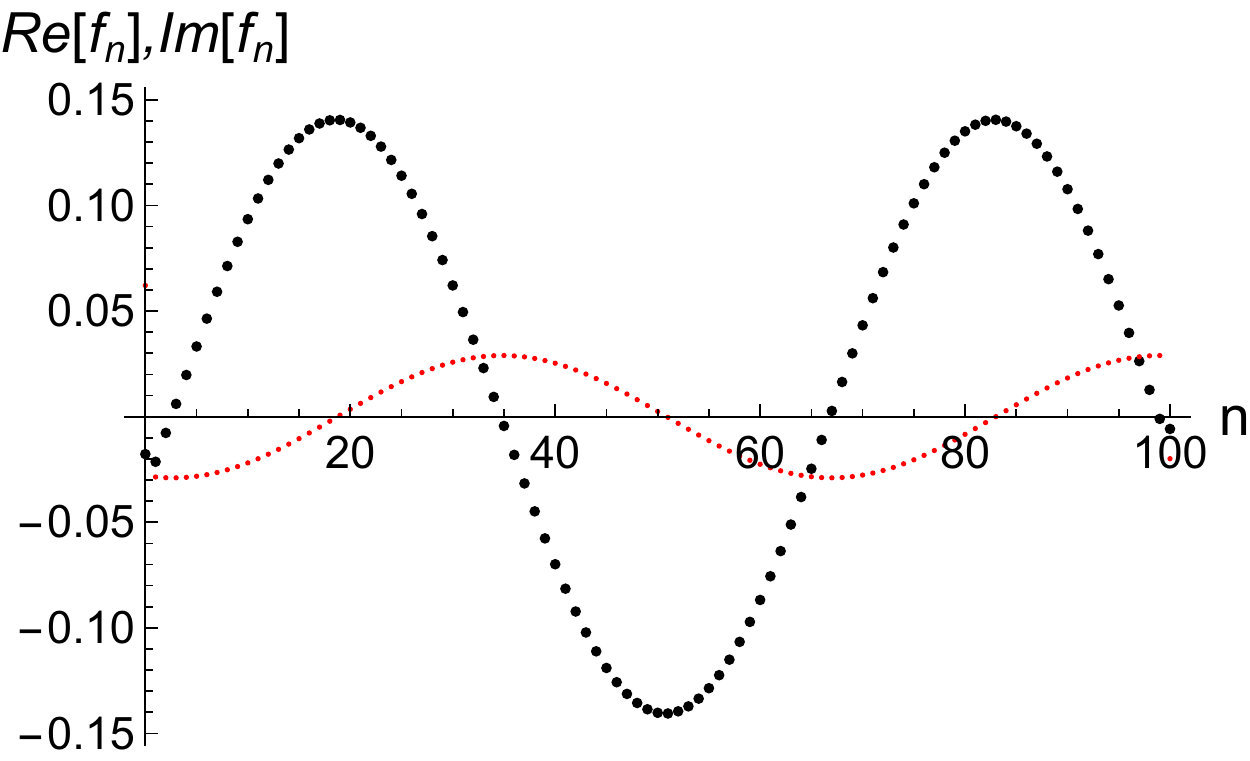}
}
\centerline{
\includegraphics[width=0.5\textwidth]{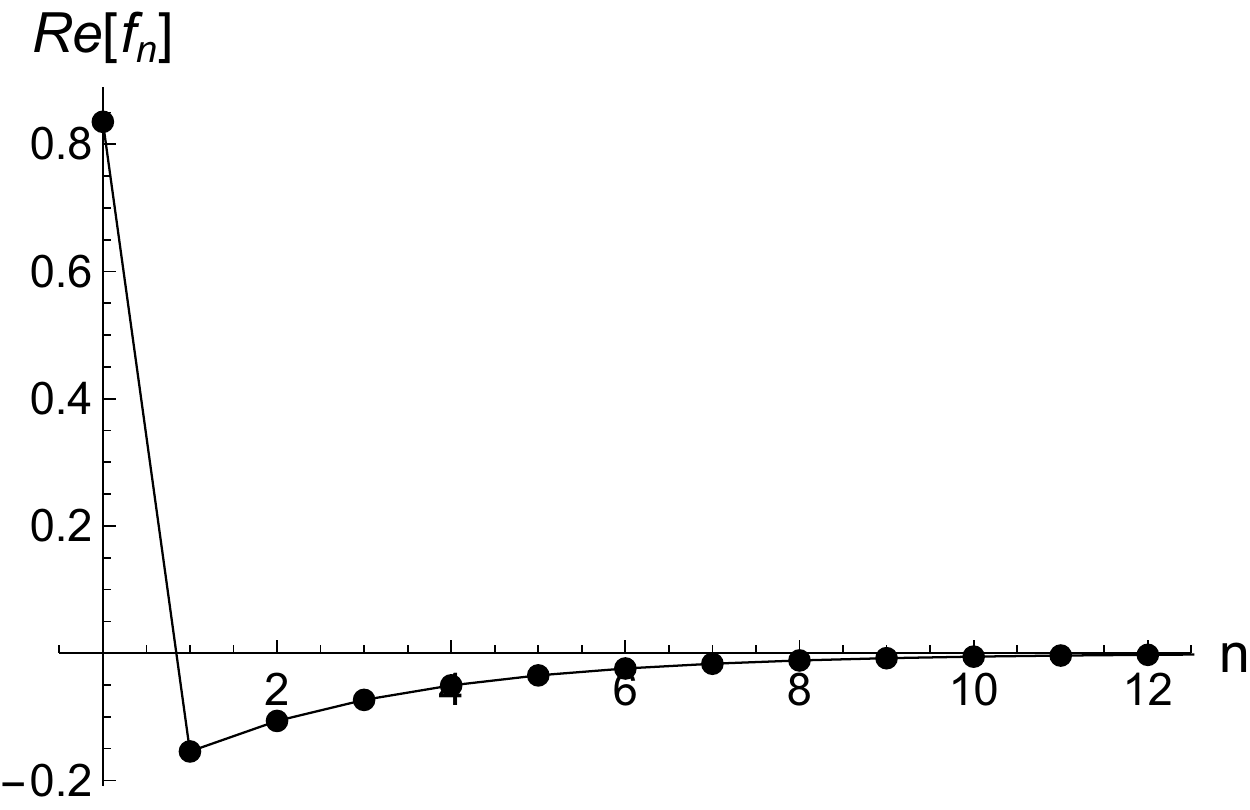}
\includegraphics[width=0.5\textwidth]{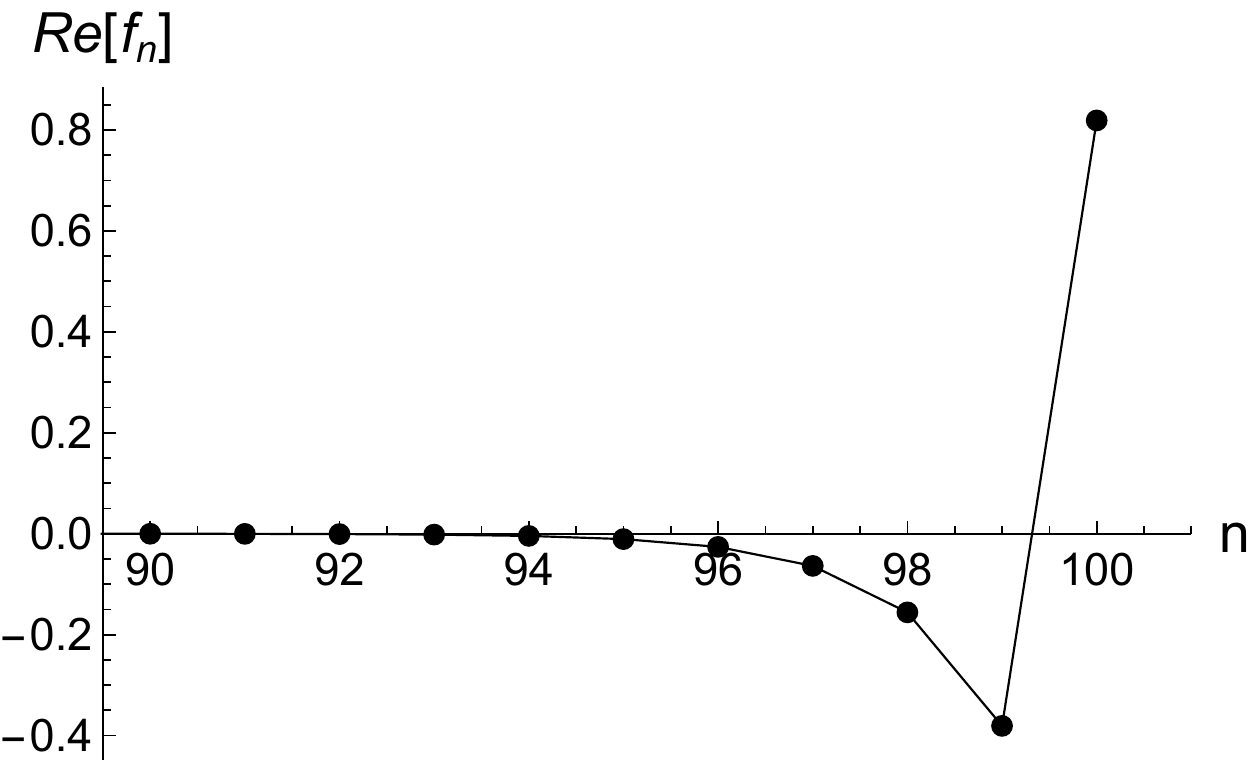}
}
\caption{Amplitudes of the Bethe vectors $f_n$ in the expansion $\langle\Psi|
  = \sum_{n=0}^{N} \bra{n} f_n $ for selected states in the
  \multiplet\ Fig.~\ref{Fig-energies}, obtained by numerical solutions of BAE
  (\ref{BAE-M1}).  \textbf{Top three panels in the left column:} Phantom Bethe
  states with the three lowest energies $E_j$ of the \multiplet, (the energy
  increases from top to bottom).  The thin lines connecting the points are
  guides for the eye. We show only ${\rm Re}[f_n]$ in the graphs.  ${\rm
    Im}[f_n]$ looks similar, with a shift by a quarter of a period as in the right
  panel.  \textbf{Top three panels in the right column:} Phantom Bethe states
  with the three highest energies $E_j$ of the real ${p_j}$ solutions (the
  energy decreases from top to bottom). Large and small points correspond to
  ${\rm Re}[f_n]$ and ${\rm Im}[f_n]$ respectively.  \textbf{Bottom row:}
  Two phantom Bethe states corresponding to the two highest energy levels in
  the \multiplet, corresponding to {imaginary} ${\rp_j}$ solutions. The
  largest (the second largest) energy of the \multiplet\ corresponds to the
  right (left) Panel in the bottom row.  }
\label{Fig-knots}
\end{figure}

\textbf{Amplitudes of phantom Bethe vectors in the \multiplet: Standing waves
  structure of the coefficients $f_n$. } It can be shown that the distribution
of the shock amplitudes $f_n$ near the lower part of the energy spectrum
(inside the \multiplet) obeys ${\rm Re}[f_n] \approx (-1)^n\sin(\frac{m \pi}
{N}) $, $m=1,2,3, \ldots$ valid for $m/N \ll 1$.  Likewise, the shock
amplitudes $f_n$ in the upper part of the energy spectrum (inside the
\multiplet\ apart from the states with exponentially decaying amplitudes $f_n$ )
obey ${\rm Re}[f_n] \approx \sin (\frac{m \pi }{N}) $, $m=1,2,3, \ldots$ where
$m/N \ll 1$ respectively.  The functions $\{f_n\}$ thus have the form of
discrete standing waves, with knots near the ``edges" $n=0,N$, see
Fig.~\ref{Fig-knots}. On the contrary, imaginary solutions ${\rp_j}$
correspond to exponentially decaying amplitudes $f_n$, see
Fig.~\ref{Fig-knots}.

 \textbf{Chirality  of  phantom Bethe vectors in the \multiplet: high current and 
modulations in the density profile }

Before calculating explicit expressions for the magnetization current $ j^z$
by using the explicit form of the Bethe function (\ref{PsiM1}), let us make
a rough estimate. The basis of the invariant subspace (\ref{basis_1}) consists
of $N+1$ basis vectors, $\{\bra{n}\}_{n=0}^N$.  The two states $\bra{0}, \bra{N}$ are pure SHS.
%the qubit phase  
%$F_k=\arg[ \bra{\phi_n(q) } \si_n^{+} \ket{\phi_1(q) }]$
% at site $k$ decreases gradually from site to site.
%$F_{k+1}-F_{k}= -  \ga$ for all links. 
The expectation value of the spin current operator $\bra{n}\,\mathbf
j_l^{z}\ket{n}$ for $n=0,N$ does not depend on the site and is given by
(\ref{jzSHS}), or, in terms of boundary parameters, by
\begin{align}
j_z^{(0)}=\frac{2\,\sin\gamma}{\ch^2(\be_+)}\,.\label{jz0}
\end{align}
The remaining basis states $\bra{n}$, i.e.~$\bra{1},\bra{2}\ldots \bra{N-1}$  have a kink in the phase at the link $n,n+1$,
where the qubit phase difference flips sign. 
%$F_{n+1}-F_{n}=  \ga$.
The expectation value of the spin operator $\bra{n}\,\mathbf j_l^{z}\ket{n}$
with  $\mathbf j_l^{z} = 2(\si_l^x \si_{l+1}^y -  \si_l^y \si_{l+1}^x)$ depends  on $l$: we find 
$\bra{n}\,\mathbf j_l^{z}\ket{n}= j_z^{(0)}$ for all links, except for the link $l,l+1$ with kink: 
$\bra{l}\,\mathbf j_l^{z}\ket{l}= -j_z^{(0)}$.
The local  current  $\bra{n}\,\mathbf j_l^{z}\ket{n}$  averaged over all links is given by 
\begin{align}
&\frac{1}{N-1} \sum_{l=1}^{N-1} \bra{n} \,\mathbf j_l^{z} \ket{n} = j_z^{(0)} \left(1-\frac{2}{N-1}\right), \quad n=1,\ldots N-1,\\
&\frac{1}{N-1} \sum_{l=1}^{N-1} \bra{n}\,\mathbf j_l^{z} \ket{n} = j_z^{(0)}, \quad n=0,N.
\end{align}
Consequently, the local current averaged over all links and all the basis states of the $G_1^{+}$ \multiplet\   is 
\begin{align}
&\langle \,\mathbf j_{local}^{z} \rangle_{G_1^+}=\frac{1}{(N+1)(N-1)} 
  \sum_{l=1}^{N-1} \sum_{n=0}^{N} \bra{n}\,\mathbf j_l^{z} \ket{n} = j_z^{(0)} \left( 1- \frac{2}{N+1} \right). \label{jz-naive}
\end{align}
 Analogously to (\ref{jz-naive}), for phantom Bethe states with two regular Bethe roots $M=2$ we obtain 
\begin{align}
&\langle\,\mathbf j_{local}^{z} \rangle_{G_2^{+}}=\frac{1}{(N-1)\dim G_2^{+}} 
  \sum_{l=1}^{N-1} \sum_{m=1}^{\dim G_2^{+}} \bra{m}\, \mathbf j_l^{z} \ket{m} = j_z^{(0)} \left( 1- \frac{4}{N+1+\frac{2}{N}} \right) 
= j_z^{(0)} \left( 1- \frac{4}{N+1}\right)+O\left( \frac{1}{N^2} \right),
\label{jz-naiveM2}
\end{align}
(here  $\ket{m}$ numerate  basis states spanning  $G_2^{+}$ ),
and so on. 

 The quantity   $\langle\, \mathbf j_{local}^{z} \rangle_{G_1^{+}}$ from
 (\ref{jz-naive}) can be regarded as a rough estimate for a typical current of
 the \multiplet; it cannot be precise since we made the equal amplitude
 assumption $f_n \equiv 1$ and in addition ignored the non-orthogonality of
 the basis states $\braket{n}{m}\neq \de_{nm}$.  However, it renders our idea
 that typical currents in the \multiplet\ can differ from the SHS current $j_z^{(0)}$  at most by $O(1/N)$ corrections, which are strictly negative, hence \textit{decreasing} its amplitude. 
Moreover, the calculations performed for special boundary
 parameters, confirm the estimate (\ref{jz-naive}) even quantitatively, see
 Eq.(\ref{CurrAverageM1}).

Using similar arguments for the magnetization profile, we conjecture that
typical transversal magnetization components must be quasi-periodic. In fact
this conjecture is confirmed by numerical simulations, see
Fig.~\ref{Fig-profiles}.

Summarizing, remarkable qualitative chirality features of the SHS are
conserved if a regular excitation is added ($M=1$). Quantitatively, they get
only slightly distorted, with degree of distortion that can be quantified.

%\subsubsection{Calculating observables: spin current and magnetization profile  for the multiplet}
Thus, all states of the \multiplet\ have distinct chiral features:
quasi-periodicity of a magnetization profile and large magnetization
current. This is especially evident for imaginary ${\rp_j}$ solutions, if the
system admits any: in fact, in this case the major contribution to the Bethe
state is given by either the pure chiral SHS $\langle 0 |$ or the SHS $\langle
N |$, while the other states contribute with exponentially decaying
amplitudes $f_n$.

Let us illustrate the calculation of a physical observable for phantom Bethe
states belonging to
the \multiplet\ at the example of the special case (\textbf{\romannumeral3})
in the previous section.
For a hermititian  Hamiltonian in the easy plane regime (see
Eq. (\ref{hermiticity_1})) we can satisfy the constraints $a_\pm\,b_\pm=1$
with the following choice of the  boundary parameters, without losing
generality:
\bee
\be_+=\be_-=0,\quad \al_\pm=-\ir\gamma+\ir\frac{\pi}{2}\quad \mbox{mod}\,\, 2\pi\ir,\quad \th_{-} - \th_{+} = \ir(N-1)\gamma \mod 2\pi \ir.
\eee
In this special case, there are two SHS: $\langle 0|$ and $\langle N|$ and the current in these two states can be calculated exactly as
\bee
j^z=\frac{\langle 0|\,\mathbf{j}_l^z|0\rangle}{\langle 0|0\rangle}=\frac{\langle N|\,\mathbf{j}_l^z|N\rangle}{\langle N|N\rangle}=2\sin\gamma.
\eee
For the remaining $N-1$ states, we know the Bethe roots $p_m = \pi m /N, 
\,m=1,\ldots,N\!-\!1$ from (\ref{BAE-iii}).
After some tedious calculations, we get the norm of the phantom Bethe vectors
\begin{align}
\langle\Psi|\Psi\rangle&=2^N\sum_{0\leq n_1,n_2\leq N}f_{n_1}f_{n_2}\Delta^{\abs{n_1-n_2}}=\frac{2^{N-1}N(1-\Delta^2)}{(1+\Delta^2-2\Delta\cos p_m)},\quad \Delta=\cos\gamma.
\end{align}
The explicit expression of the current is 
\begin{align}
{j}^z(m)&=\frac{\langle \Psi|\,\mathbf{j}_l^z|\Psi\rangle}{\langle \Psi|\Psi\rangle}=2\sin\gamma\,\frac{\sum_{0\leq n_1,n_2\leq N}f_{n_1}f_{n_2}\Delta^{\abs{n_1-n_2}}+2f_0^2-2f_0^2\Delta^{-2}}{\sum_{0\leq n_1,n_2\leq N}f_{n_1}f_{n_2}\Delta^{\abs{n_1-n_2}}}\no\\
&=2\sin\gamma\left(1-\frac{4(1-\cos^2 p_m)}{N(1+\Delta^2-2\Delta\cos p_m)}\right),\quad p_m=\frac{\pi m}{N},\quad m=1,\ldots,N-1,\label{current_case(iii)}
\end{align}
where we emphasize the dependence of the current on the particular state by
using the state's ordinal number as argument in ${j}^z(m)$.  The expression
(\ref{current_case(iii)}) can also be applied to the other two SHS letting
$m=0,N$. Suppose $0<\gamma<\pi$. The respective SHS current $j^z(0)=j^z(N)$ is
maximal and the minimal value is among $j^z(m)$ and $j^z(m\!+\!1)$ with
$\frac{m\pi}{N}\leq \gamma\leq \frac{(m+1)\pi}{N}$. In the case $N\gg 1$
\bee
\min_{m} j^z(m)=j_z^{(0)} \left(1-\frac{4 }{N}\right)+o\left(\frac{1}{N}\right).\label{MinimalCurrentM1}
\eee
We can calculate the average current of the \multiplet, 
\begin{align}
\langle\, \mathbf j^z \rangle_{G_1^{+}} = \frac{1}{N+1} \sum_{m=0}^{N} j^z(m)=j_z^{(0)} \left(1-\frac{2 }{N+1} \right)+O\left( \frac{1}{N^2}\right),
\label{CurrAverageM1}
\end{align}
in qualitative accordance with our naive estimate (\ref{jz-naive}).
The fact that corrections to the current are strictly negative originates from
the influence of the kinks in the states as explained in the paragraph
following (\ref{jz0}). The
links with kinks sustain local current of opposite sign, reducing
the  current amplitude.  The invariant subspace for $M=0$ contains one state (SHS)
with no kinks and the respective SHS current $j_z^{(0)}$ is
maximal. The $N+1$-dimensional invariant subspace for $M=1$ consists of states
with $0$ or $1$ kink and the average current reduces by the fraction $\frac{2
}{N+1}$, see (\ref{CurrAverageM1}), (\ref{jz-naive}).  For $M=2$, the
invariant subspace consists of states with $0,1$ or $2$ kinks, leading to
a further decrease of the average current, as predicted by Eq.(\ref{jz-naiveM2}).

We conclude that the inclusion of further
regular Bethe roots (in case of larger $M$) makes the chiral properties of the
phantom Bethe states less pronounced. However the average
multiplet current can decrease significantly, only if typical multiplet
basis states contain sizable proportions of kinks, meaning $M/N = O(1)$.
The accurate analysis of the quantity (\ref{CurrAverageM1}) for arbitrary $M$
requires further investigation and is out of the present scope.

A high average current is not the only chiral feature of phantom Bethe
states. Another typical feature is the large periodic modulation of the
magnetization profile. We find that inclusions of regular Bethe roots distort
the perfectly periodic spin helix structure. The degree of distortion naturally
depends on the number $M$ of regular Bethe roots involved. If $M/N\ll 1$,
modulations of the magnetization profile are clearly visible for all members
of the multiplet.  We show typical magnetization profiles in
Fig.~\ref{Fig-profiles} for $M=1$.

Our analytic results are fully confirmed by numerical simulations, done for
large system size $N$.  In Fig.~\ref{Fig-knots} we show typical amplitudes
of phantom Bethe vectors for $M=1$.  In Fig.~\ref{Fig-profiles} we
show typical magnetization profiles.

\begin{figure}[tbp]
\centerline{
\includegraphics[width=0.48\textwidth]{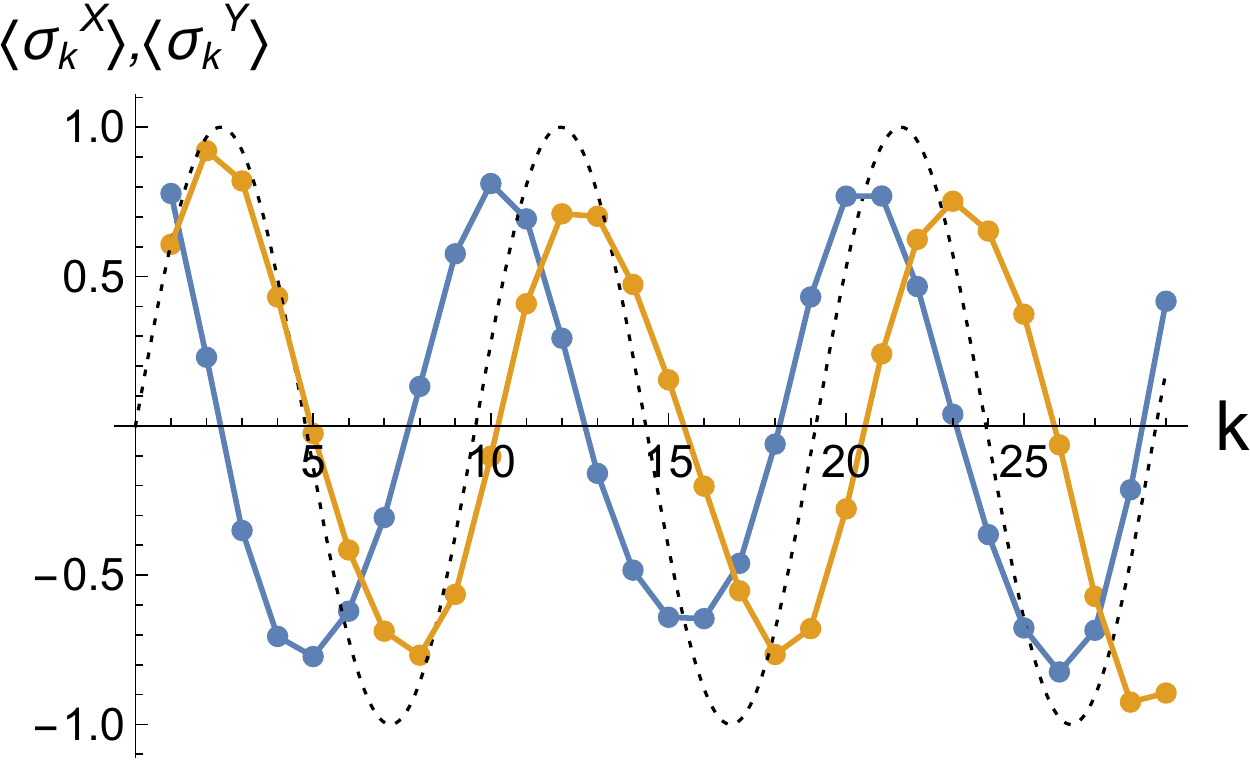}
}
\caption{Transversal components of the magnetization profiles $\langle \si_k^x
  \rangle, \langle \si_k^y \rangle $ (blue and yellow curves, respectively),
  for a typical phantom Bethe state from $G_1^{+}$ (one regular Bethe root),
  versus site number $k$.   Parameters: $N=29$, $\eta= \ir \ga = \ir \frac {103 \pi}{493}  $,
  $\De=\cos \frac {103 \pi}{493} \approx 0.79$, 
$\al_{\pm}= -\eta,\,\, \, \be_\pm=0, \,\,\,\th_-= \ir \pi, \ \th_+ = \ir(\pi - \Phi)$,
and $\Phi =( N-1 ) \vfi$.
The energy of the chosen state is $E_m= -5.39$ and the corresponding current
is $j_z=1.216$, close to the asymptotic SHS current $j_{SHS} \approx
  1.22$.  The dotted line shows the $y$-component of an ``ideal chiral" SHS profile
  for the same value of the anisotropy and is given for comparison. }
\label{Fig-profiles}
\end{figure}

\section{Discussion}

We have analyzed the integrable XXZ Heisenberg spin chain with open
  boundary conditions and have described a novel type of solutions to the
  Bethe Ansatz equations containing phantom (infinite) Bethe roots, as well as
  regular (finite) Bethe roots.  These solutions appear under condition
  (\ref{ConditionPhantom}) which leads to a complete decoupling of the Bethe
  Ansatz equations for phantom and regular Bethe roots.  Phantom Bethe roots do not
  contribute to the energy of the system, which in case of spin chains with
  periodic boundary condition leads to degeneracies of the energies
  \cite{PhantomShort} which we refer to as phantom excitations.

Condition (\ref{ConditionPhantom}) has appeared in
  \cite{OffDiagonal03,Nepomechie2003,Rafael2003,Cao2013off} as a technical
  condition for the applicability of a modified Algebraic Bethe Ansatz,
  based on special properties of Sklyanin's $K$-matrices. In the present
  manuscript, we have unveiled its meaning as a condition for the splitting of
  the Hilbert space into two invariant chiral subspaces $G_M^{\pm}$.  
In addition, here we  used phantom Bethe roots  as a useful shortcut to obtain  reduced BAEs (18), (21).
The
  integer parameter $0\leq M<N$ entering (\ref{ConditionPhantom}) determines
  the dimensions of the invariant subspaces, $\dim G_M^{+}= \sum_{k=0}^M
  \binom{N}{k}$ and $\dim G_M^{-}=2^N-\dim G_M^{+}$.

The meaning of the key parameter $M$, and its dual $\widetilde {M}=N-M-1$,
parametrizing the hyperplane (\ref{ConditionPhantom}) is twofold. On one hand,
$M$ is the maximal number of kinks in the chiral basis vectors
(\ref{ResShockPlus}) of $G_M^{+}$. Analogously, $\widetilde {M}$ is the
maximal number of kinks in the chiral basis vectors of $G_M^{-}$.  On the
other hand, the fulfillment of (\ref{ConditionPhantom}) entails a reduction of
the number of regular Bethe roots from $N$ in the general ``inhomogeneous''
BAE to $M$ or $\widetilde {M}$ in the reduced sets of BAE. The solutions of
the two BAE sets give all Bethe eigenstates, which necessarily possess chiral
properties since the nature of the invariant subspaces is chiral.

Conversely, we understand condition (\ref{ConditionPhantom}) as a criterion
for the occurrence of phantom Bethe roots \cite{PhantomShort} among the $N$
initial Bethe roots: the phantom Bethe roots decouple from the general
``inhomogeneous'' BAE, leading to the two (dual) homogeneous BAE sets for $M$
or $\widetilde {M}$ regular Bethe roots. The BAE for the phantom
Bethe roots are not trivially satisfied just by these roots lying at infinity. The
equations require a specific arrangement of the roots in a string that is
unrelated to the usual TBA strings.

The appearance of invariant subspaces and the splitting of the set of
eigenvectors into blocks is sowewhat similar to the occurrence of invariant
subspaces for the periodic XXZ spin chain with fixed values of
magnetization. There are however crucial differences: in the periodic case
there are $N+1$ blocks, with magnetizations $-N/2, -N/2+1, ..., N/2$ and BAE
with a number of Bethe roots specific for each block, ranging
from 0 to $N/2$ for even $N$ and to $(N-1)/2$ for odd $N$.

In contrast, in the open XXZ model fulfilling
(\ref{ConditionPhantom}) with $0\leq M<N$ all eigenstates split into just two
blocks, with basis vectors that are chiral, which leads to two sets of BAE
with a total number of $M$ roots or $\widetilde M$ roots.  The latter fact
leads to highly unusual properties of the respective eigenstates such as high
magnetization currents and quasiperiodic magnetization profiles.

The main result of this paper is the proof that in the open integrable XXZ
spin chain, the occurrence of phantom Bethe roots entails the splitting of
the Hilbert space into two chiral invariant subspaces.  Based on the
splitting, we are able to construct explicit Bethe vectors, the eigenstates of
the open XXZ model with fine-tuned non-diagonal boundary fields, and
investigate their properties to the same degree of detail as for the
periodic spin chain with $U(1)$ symmetry for a given magnetization sector. The
phantom Bethe eigenstates in open systems are very unusual and carry distinct
chiral properties. This is due to the underlying chiral nature of the basis
states, constituting the respective invariant subspaces. Our results can be
used for the generation of stable spin helix states in experimental setups
allowing to realize a paradigmatic XXZ model with tunable anisotropy
\cite{2020NatureSpinHelix} as argued in \cite{PhantomShort}.
 
It would be interesting to extend our results to other integrable systems with
phantom Bethe roots e.g.~to the spin-1 Fateev-Zamolodchikov model with open
boundary conditions, see \cite{PhantomShort}.

\section{acknowledgments}
 Financial support from the
  Deutsche Forschungsgemeinschaft through DFG project KL 645/20-1,
is gratefully   acknowledged. 
Xin Zhang thanks the Alexander von Humboldt Foundation for  financial support. Xin Zhang would like to thank Yupeng Wang and Junpeng Cao for discussions.
%\end{acknowledgments}

\bibliographystyle{IOP}
%\bibliography{Reference_XXZ}

\newpage

\appendix

\section{Proof of Completeness}\label{Proof of Completeness}
Here we prove that the basis of states spanning  $G_M^{+}$ and $ G_M^{-}$ is complete, i.e. 
\begin{align}
&{\cal H} = G_M^{+} \oplus G_M^{-}.\label{Completeness}
\end{align}
where ${\cal H}$ is the full Hilbert space. First we prove, that the
vectors $\langle\,{g_{+,k}}| \in G_M^{+}$, $k=1,2,\ldots,\dim G_M^{+}$ are all
independent for $\eta \neq 0,\ir\pi$ (for the cases $\eta= 0,\ir\pi$, a
jump $\phi_n(x) \rightarrow \phi_n(x+1)\equiv \phi_n(x)$ is trivial). We
restrict our proof to the most ``unfavourable", extreme case of
$\eta=\ir\pi/2$, $\be_{-}=0$ under the hermiticity condition
(\ref{hermiticity_1}). For this special case, the property
$\phi_n(x)\phi_n^\dagger(x+1)\!=\!0$ renders all states $\bra{\,g_{+,k}} \in
G_M^{+}$ being pairwise orthonormal, $\braket{\,g_{+,k} }{g_{+,l}}= 2^N\de_{kl}
\braket{\,g_{+,k} }{g_{+,k}} $. Thus, even in the most ``unfavourable" setting
all $\bra{\,g_{+,k}} $ are independent, and the same is valid for the basis
vectors $|{g_{-,k}}\rangle\!\rangle $ spanning $G_M^{-}$.

In the next step, we return to the general setup and show that any two
vectors $g_{-} \in G_M^{-}$ and $g_{+} \in G_M^{+}$ are orthogonal. Define the
function $y(n,j_n,k_n)$ as
\bee
\phi_{n}(j_n)\,\tilde{\phi}_n(k_n)=%1-\eE^{2(N-n-j_n-k_n)}=%
1-\eE^{2y(n,j_n,k_n)\eta},\quad y(n,j_n,k_n)\!=\!j_n+k_n-n+1.\label{def;y}
\eee
When $y(n,j_n,k_n)=0$, the local vectors $\phi_{n}(j_n)$ and $\tilde{\phi}_n(k_n)$ are orthogonal.
From the definition of (\ref{ResShockPlus}) and (\ref{ResShockMinus}), any basis vector belonging to $G_M^+$ is a tensor product of $\phi_{n}(j_n)$ with $0\leq j_1\leq j_{2}\ldots \leq j_N\leq M$ and any basis vector belonging to $G_M^-$ is a tensor product of $\tilde\phi_{n}(k_n)$ with $0\leq k_1\leq k_{2}\ldots\leq k_N\leq \widetilde M$. It is easy to find
\bee
&&y(n+1,j_{n+1},k_{n+1})-y(n,j_n,k_n)=0,\pm 1,\no\\
&&y(N,j_N,k_N)\leq 0,\quad y(1,j_1,k_1)\geq 0.\label{orthogonality}
\eee 
so that $y(n,j_n,k_n)=0$ holds at least for one point $n$ ($1\leq n\leq
N$). It shows that any pair  of vectors $g_{-} \in  G_M^{-}$  and  $g_{+} \in
G_M^{+}$ is orthogonal.

The dimension of $G_M^{+}$ is equal to the total number of different tuples 
$(n_1,\ldots ,n_k)$  with $1\leq n_1<n_2<\ldots <n_k \leq N$ over all $k=0,
\ldots M$, which is given by
\begin{align}
&\dim G_M^{+} =  \sum_{m=0}^M \binom{N}{m}= \binom{N}{0} + \binom{N}{1} + \ldots \binom{N}{M}\,.
\end{align}

Analogously, the dimension of $G_M^{-}$ is 
\begin{align}
&\dim G_{M}^{-} = \sum_{m=0}^{\widetilde M}  \binom{N}{m}=  \sum_{n=M+1}^{N}  \binom{N}{n}\,.
\end{align}
The sum of dimensions $\dim G_{M}^{+} +\dim G_M^{-}=  \sum_{n=0}^{N}
\binom{N}{n}= 2^N$ is identical to the dimension of the
 total Hilbert space  ${\cal H}$. Hence the basis vectors in (\ref{ResShockPlus}) and (\ref{ResShockMinus}) span  $G_M^{+}$ and $G_M^-$ respectively, hence 
(\ref{Completeness}) is proved.

\section{The invariance property of $G_M^\pm$}

Here we prove the  theorem  for arbitrary Hamiltonian $H$ of type
(\ref{Hamiltonian}), satisfying (\ref{ConditionPhantomPlus}), whether
being hermitian or not.
It is easy to prove that
\bee 
&&\phi_n(x)\phi_{n+1}(x)h_{n,n+1}=\sh\eta\,\phi_n(x)\,\sigma_n^z\,\phi_{n+1}(x)-\sh\eta\,\phi_n(x)\,\phi_{n+1}(x)\,\sigma_{n+1}^z,\label{bulk_1}\\[2pt]
&&\phi_n(x\!-\!1)\phi_{n+1}(x)h_{n,n+1}=\sh\eta\,\phi_n(x\!-\!1)\,\phi_{n+1}(x)\,\sigma_{n+1}^z-\sh\eta\,\phi_n(x\!-\!1)\,\sigma_n^z\,\phi_{n+1}(x),\label{bulk_2}\\[2pt]
&&\phi_1(x)h_1=-\frac{\sh\eta\,\ch(\al_-\!+\!\be_-\!+\!2x\eta)}{\sh(\al_-)\ch(\be_-)}\phi_1(x)\no\\[2pt]
&&+\frac{\sh\eta}{\sh(\al_-)\ch(\be_-)}\left(\ch(\al_-)\sh(\be_-)-\sh(\al_-\!+\!\be_-\!+\!2x\eta)\right)\phi_1(x)\,\sigma_1^z,\label{left}\\[2pt]
&&\phi_N(x)h_N=-\frac{\sh\eta\,\ch(\al_+\!+\!\be_+\!+\!2(M\!-\!x)\eta)}{\sh(\al_+)\ch(\be_+)}\phi_N(x)\no\\[2pt]
&&+\frac{\sh\eta}{\sh(\al_+)\ch(\be_+)}(\sh(\al_+\!+\!\be_+\!+\!2(M\!-\!x)\eta)-\ch(\al_+)\sh(\be_+))\phi_N(x)\,\sigma_N^z.\label{right}
\eee
%Here we used the notation \bee &&x^+=x+2,\qquad x^{++}=x+4,\quad \rho_\pm=\al_\pm+\be_\pm.\label{rho}\eee
From (\ref{bulk_1}) - (\ref{right}), it is obvious that
\bee
\Phi_+(n_1,\ldots,n_k)\,H= C_0(n_1,\ldots,n_k)\Phi_+(n_1,\ldots,n_k)+\sum_{n=1}^NC_n(n_1,\ldots,n_k)\Phi_+(n_1,\ldots,n_k)\,\sigma_n^z,\label{Phi_H}
\eee 
where $C_{n}(n_1,\ldots,n_k),\,\,\,n=0,1,\ldots,N$, are some constants. Using Eqs. (\ref{bulk_1}) - (\ref{right}) repeatedly and checking the detailed coefficients, we can prove that the coefficient $C_{n}(n_1,\ldots,n_k)$
is zero when $\Phi_+(n_1,\ldots,n_k)$ has some special structure which is
realized under any of the following conditions
\begin{align}
\begin{aligned}\label{C;property}
&\Phi_+(n_1,\ldots,n_M)=\phi_1(0)\phi_2(0)\cdots,\,\,\mbox{i.e.}\,\, 2\!\leq\!n_1\!<\!n_2\ldots\!<\!n_M\!\leq\! N,\qquad C_{1}(n_1,\ldots,n_M)=0,\\
&\Phi_+(n_1,\ldots,n_k)=\cdots\phi_{N-1}(M)\phi_N(M),\,\,\,\mbox{i.e.}\,\, 1\!\leq\!n_1\!<\!n_2\ldots\!<\!n_k\!\leq\! N\!-\!2,\qquad\quad  C_{N}(n_1,\ldots,n_k)=0,\\
& \Phi_+(n_1,\ldots,n_k)=\cdots \phi_{n-1}(x)\phi_n(x)\phi_{n+1}(x)\cdots,\,\,\mbox{i.e.}\,\, n_m\!\leq\!n\!-\!2,\,\,n_{m+1}\!\geq\!n\!+\!1,\!\quad C_{n}(n_1,\ldots,n_k)=0,\\
&\Phi_+(n_1,\ldots,n_k)=\cdots \phi_{n-1}(x-1)\phi_n(x)\phi_{n+1}(x+1)\cdots,\,\,\mbox{i.e.}\,\, n_m\!=\!n\!-\!1,\,\,n_{m+1}\!=\!n,\!\quad C_{n}(n_1,\ldots,n_k)=0.
\end{aligned}
\end{align}
We have two useful identities
\bee 
\begin{aligned}\label{psi;phi}
&\phi_n(x)\,\sigma_n^z=-\frac{\ch\eta}{\sh\eta}\phi_n(x)+\frac{\eE^\eta}{\sh\eta}\phi_n(x-1),\\
&\phi_n(x-1)\,\sigma_n^z=\frac{\ch\eta}{\sh\eta}\phi_n(x-1)-\frac{\eE^{-\eta}}{\sh\eta}\phi_n(x).
\end{aligned}
\eee
Then, we obtain 
\bee
\begin{aligned}\label{psi;phi2}
&\phi_{n}(x-1)\phi_{n+1}(x)\sigma_n^z=\frac{\ch\eta}{\sh\eta}\phi_n(x-1)\phi_{n+1}(x)-\frac{\eE^{-\eta}}{\sh\eta}\phi_n(x)\phi_{n+1}(x),\\
&\phi_{n}(x)\phi_{n+1}(x)\sigma_n^z=-\frac{\ch\eta}{\sh\eta}\phi_n(x)\phi_{n+1}(x)+\frac{\eE^{\eta}}{\sh\eta}\phi_n(x-1)\phi_{n+1}(x),\\
&\phi_{n-1}(x)\phi_n(x+1)\sigma_n^z=-\frac{\ch\eta}{\sh\eta}\phi_{n-1}(x)\phi_n(x+1)+\frac{\eE^\eta}{\sh\eta}\phi_{n-1}(x)\phi_n(x),\\
&\phi_{n-1}(x)\phi_n(x)\sigma_n^z=\frac{\ch\eta}{\sh\eta}\phi_{n-1}(x)\phi_n(x)-\frac{\eE^{-\eta}}{\sh\eta}\phi_{n-1}(x)\phi_n(x+1).
\end{aligned}
\eee
Let us extend the definition of $\Phi_+(n_1,\ldots,n_k)$ as 
\bee
\Phi_+(0,n_1,\ldots,n_k)\equiv\Phi_+(n_1,\ldots,n_k),\quad1\leq n_1<\ldots<n_k\leq N.\label{extention}
\eee
Using Eqs.~(\ref{Phi_H}), (\ref{psi;phi2}) and the notation (\ref{extention}),
we can prove that
$\Phi_+(n_1,\ldots,n_k)\,H$ for $(n_{k-1},n_k)\!\neq\! (N-1,N)$ is a linear combination of $$\Phi_+(n_1,\ldots,n_k)\,\,\mbox{and} \,\, \,\Phi_+(n_1,\ldots,n_m\pm 1,\ldots,n_k),\,\,m=1,\ldots,k.$$ 
Here the actions $\Phi_+(n_1\!=\!0,n_2\ldots,n_k)H\!\to\! \Phi_+(n_1\!=\!1,n_2\ldots,n_k)$ and $\Phi_+(n_1\!=\!1,n_2\ldots,n_k)H\!\to\! \Phi_+(n_1\!=\!0,n_2\ldots,n_k)$ represent the generation and annihilation of a kink at the left boundary respectively.
Due to Eqs. (\ref{C;property}) and (\ref{psi;phi2}), the following unwanted structures will not appear  
\begin{align}
&\Phi_+(\ldots,n_j,n_{j+1}=n_j,\,\ldots),\no\\
&\Phi_+(n_1,\ldots,n_{M+1}),\quad 1\leq n_1<n_2<\ldots<n_{M+1}\leq N.
\end{align}
In words, the Hamiltonian $H$ acting on $
\Phi_+(n_1,\ldots,n_k)$ with $(n_{k-1},n_k)\neq (N-1,N)$ can not degenerate any state beyond our basis vectors $\{\Phi_+(\ldots)\}$. 
Some nontrivial situations arise when $H$ acts on $\Phi_+(n_1,\ldots,n_{k-1}\!=\!N\!-\!1,n_k\!=\!N)$
\bee
\Phi_+(\ldots,N\!-\!1,N)H=\cdots+F(\ldots,N\!-\!1,N)\,\Phi_+(\ldots,N,N),\label{unwanted_term}
\eee
where $\cdots$ on the RHS of (\ref{unwanted_term}) denotes a linear combination of some vectors belonging to $\Phi_+(\ldots)$ in (\ref{ResShockPlus}) and $F(\ldots,N\!-\!1,N)$ is a constant. These ``unwanted'' vectors $\left\{\Phi_+(\ldots,N,N)\right\}$ are  defined by replacing $\phi_{N-1}(M\!-\!2)\phi_{N}(M\!-\!1)$ in $\Phi_+(\ldots,N\!-\!1,N)$ with $\phi_{N-1}(M\!-\!2)\,\phi_{N}(M\!-\!2)$.
In fact $\Phi_+(\ldots,N,N)$ is not an extra independent vector. Following the method in (\ref{def;y}) and (\ref{orthogonality}) we can prove $\Phi_+(\ldots,N,N)$ is orthogonal to all the basis vectors in $G_M^-$ and can be expanded in $G_M^{+}$ basis. Thus  the invariance property of the subspace $G_M^{+}$ is proved. 
Analogously,  we  prove the invariance property of the set $G_M^{-}$. 

\bigskip 
\section{The proof of Eqs.~(\ref{M1;b1})-(\ref{M1;b5})}

Using Eqs.~(\ref{bulk_1}) - (\ref{right}) and (\ref{psi;phi}), we find
\bee 
&&\langle 0|H=\left(\frac{\sh\eta\,\ch(\al_-\!+\!\be_-)}{\sh(\al_-)\ch(\be_-)}-\frac{\sh\eta\ch(\al_+\!+\!\be_+)}{\sh(\al_+)\ch(\be_+)}\right)\langle 0|\no\\
&&\hspace{1.6cm}-\frac{2\sh\eta\,\ch(\al_-\!+\!\be_-\!+\!\eta)}{\sh(\al_-)\ch(\be_-)}\langle 1|,\label{H0}\\[2pt]
&&\langle N|H=\left(\frac{\sh\eta\,\ch(\al_+\!+\!\be_+)}{\sh(\al_+)\ch(\be_+)}-\frac{\sh\eta\,\ch(\al_-\!+\!\be_-)}{\sh(\al_-)\ch(\be_-)}\right)\langle N|\no\\
&&\hspace{1.6cm}-\frac{2\sh\eta\,\ch(\al_+\!+\!\be_+\!+\!\eta)}{\sh(\al_+)\ch(\be_+)}\langle N\!-\!1|,\label{HN}\\[2pt]
&&\langle n|H=-\left(\frac{\sh\eta\ch(\al_-\!+\!\be_-)}{\sh(\al_-)\ch(\be_-)}+\frac{\sh\eta\ch(\al_+\!+\!\be_+)}{\sh(\al_+)\ch(\be_+)}+{4\ch\eta}\right)\langle n|\no\\[2pt]
&&\qquad\qquad+2\langle n\!-\!1|+2\langle n\!+\!1|,\quad n=2,\ldots,N.\label{Hn}
\eee
%where $\rho_\pm=\al_\pm+\be_\pm$. 
We have two useful identities
\begin{align}
&\frac{\sh\eta\,\ch(\al_\pm+\be_\pm)}{\sh(\al_\pm)\,\ch(\be_\pm)}={a_\pm+b_\pm}-2\ch\eta,\label{Iden-1}\\
&\frac{\sh\eta\,\ch(\al_\pm+\be_\pm+\eta)}{\sh(\al_\pm)\,\ch(\be_\pm)}={a_\pm\,b_\pm-1}.\label{Iden-2}
\end{align}
where $a_\pm,b_\pm$ are given by (\ref{def;ab}).
Using Eqs.~(\ref{Iden-1}), (\ref{Iden-2}), we can rewrite Eqs.~(\ref{H0}) - (\ref{Hn}) in terms of $a_\pm$ and $b_\pm$ and then easily obtain Eqs.~(\ref{M1;b1}) - (\ref{M1;b5}).

\section{The solutions of BAE (\ref{BAE;CBA}) with $M=1$}
Recall the BAE for the $M=1$ case
\bee \eE^{2\ir
	N{\rP}}\prod_{\sigma=\pm}\frac{a_\sigma-\,\eE^{\ir{\rP}}}{1-a_\sigma\,\eE^{\ir{\rP}}}
\,\,\frac{b_\sigma-\,\eE^{\ir{\rP}}}{1-b_\sigma\,\eE^{\ir{\rP}}}=1.\label{BAE_M1}
\eee
{When $M=1$, we can treat $\eE^{\ir{\rP}}$ as an unknown parameter which
	satisfies the unary equation (\ref{BAE_M1}) with degree $2N\!+\!4$. It is
	easy to prove that $\eE^{\ir{\rP}}\!=\!\pm 1$ are always two trivial
	solutions. Replacing $\eE^{\ir{\rP}}$ with $\eE^{-\ir{\rP}}$, the BAE still
	holds which implies that $\eE^{\ir{\rP}}$ and $\eE^{-\ir{\rP}}$ are
	equivalent (see Eq.~(\ref{Energy;CBA})). So we can summarize that the BAE
	(\ref{BAE_M1}) has $N\!+\!1$ independent valid solutions.}
If the Hamiltonian $H$ is hermitian, $\rP$ can be a real or purely imaginary
number. Here we only consider the easy plane regime case with (\ref{hermiticity_1}). 
\paragraph{Imaginary solutions}
Introduce the following auxiliary function 
\bee
&&Y(x)=x^{2N}\prod_{\sigma=\pm}\left(x-a_\sigma\right)\left(x-b_\sigma\right)-\prod_{\sigma=\pm}\left(1-a_\sigma \,x\right)\left(1\!-b_\sigma\,x\right).\label{W_function}
\eee
The zero points of $Y(x)$ correspond to the solution of BAE (\ref{BAE_M1}). Suppose $0<a_-<1<a_+$. We find that
\bee &&Y(1)=0,\quad Y(0)=-1,\quad Y\left(a_+^{-1}\right)=a_+^{-2N-4}\prod_{\sigma=\pm}(1-a_+\,a_\sigma)(1-a_+\,b_\sigma).
%&&W\left(a_+\right)=a_+^{2N}(1-a^2_+)(1-a_+a_-)(1-a_+\,b_+)(1+a_+\,b_-).
\eee
When $a_+\,a_->1$, we find $Y\left(a_+^{-1}\right)>0$ and equation $Y(x)=0$ has a solution in the interval $\left(0,\,a_+^{-1}\right)$. When $a_+\,a_-<1$, we consider the derivative  of $Y(x)$ at the point $x=1$
\bee 
&&Y'(1)=\left(\prod_{\sigma=\pm}\left(1-a_\sigma\right)\left(1\!-b_\sigma\right)\right)\left(2N-\sum_{\sigma=\pm}\frac{a_\sigma+1}{a_\sigma-1}-\sum_{\sigma=\pm}\frac{b_\sigma+1}{b_\sigma-1}\right).\qquad\quad
\eee
If $Y'(1)<0$, i.e. 
\bee 
N\!>\!\frac12\left(\sum_{\sigma=\pm}\frac{a_\sigma+1}{a_\sigma-1}+\sum_{\sigma=\pm}\frac{b_\sigma+1}{b_\sigma-1}\right),\label{N_min}
\eee 
the function $Y(x)$ has a zero in the interval $\left(a_+^{-1},\,1\right)$.

Then suppose that $1<a_-<a_+$. We can prove that
\bee 
Y(0)<0,\quad Y\left(a_+^{-1}\right)>0, \quad Y\left(a_-^{-1}\right)>0,\quad Y(1)=0.
\eee
Obviously, there exists a real solution in the interval
$\left(0,\,a_+^{-1}\right)$. If $Y'(1)>0$, i.e.~the inequality (\ref{N_min})
holds, there will be another real solution in the interval
$\left(a_-^{-1},\,1\right)$. 
When $N\gg1$ the purely imaginary solutions of (\ref{BAE_M1}) are very simple. We find  \bee 
\eE^{2\ir N{\rP}}\to \begin{cases}
	0, & -1<\eE^{\ir\rP}<1,\\
	\infty, & \eE^{\ir\rP}<-1\,\, {\mbox{or}}\,\, \eE^{\ir\rP}>1.
\end{cases}\no
\eee
So there should be a purely imaginary solution at the point $\eE^{\ir\rP}\!=\!{a_\sigma^{-1}}+O\left(\frac 1N\right)$ in case of $a_\sigma>1$ or
$a_\sigma<-1$. We can use a similar method to analyze the
distribution of real solutions in other cases.
%With the same method, we can prove that the BAE (\ref{BAE_M1}) has another real solution at the point$\eE^{\ir\rP}\!=\!b_\sigma^{-1}+O\left(\frac1N\right)$ in case of $b_\sigma>1$ or$b_\sigma<-1$.

\paragraph{Real solutions}
%When $a_l=\pm 1$, $\frac{1-a_l\,\eE^{{\rP}}}{a_l-\,\eE^{{\rP}}}\equiv \pm 1$ and the BAE will be simplified.%
Suppose that $\rP$ is a real number $\rP=\epsilon$ with $\epsilon$ being a small positive real number, then we can prove  
\bee 
&&\frac{a_\sigma-\,\eE^{\ir{\rP}}}{1-a_\sigma\,\eE^{\ir{\rP}}}=\frac{a_\sigma-(1\!+\!\ir\epsilon\!+\!\cdots)}{1-a_\sigma(1\!+\!\ir\epsilon\!+\!\cdots)}=-1+\ir\frac{a_\sigma+1}{a_\sigma-1}\epsilon+\cdots,\quad
\eee 
with $\left|\frac{\ir\epsilon}{1-a_\sigma}\right|\!\ll\! 1$. 
If
\bee 
\left|\frac{\ir\epsilon}{1-a_\sigma}\right|\!\ll\!1,\,\,\,\left|\frac{\ir\epsilon}{1-b_\sigma}\right|\!\ll\!1,\quad \sigma\!=\!\pm,\no\label{unity_root_1}
\eee
$\rP=\epsilon=\frac{m\pi}{N}+O\left(\frac{1}{N^2}\right)$ is a solution of BAE (\ref{BAE_M1}) provided that $0\!<\!\frac mN\!\ll\! 1$.

Now we suppose that $\rP$ is a real number $\rP=\pi-\epsilon$ with $\epsilon$ being a small positive real number, then we find
\bee 
&&\frac{a_\sigma-\,\eE^{\ir{\rP}}}{1-a_\sigma\,\eE^{\ir{\rP}}}=\frac{a_\sigma+(1\!-\!\ir\epsilon\!+\!\cdots)}{1+a_\sigma(1\!-\!\ir\epsilon\!+\!\cdots)}=1+\ir\frac{a_\sigma-1}{a_\sigma+1}\epsilon+\cdots,
\eee
with $\left|\frac{\ir\epsilon}{1+a_\sigma}\right|\!\ll\! 1$. If \bee 
\left|\frac{\ir\epsilon}{1+a_\sigma}\right|\!\ll\!1,\,\,\,\left|\frac{\ir\epsilon}{1+b_\sigma}\right|\!\ll\!1,\quad \sigma\!=\!\pm,\no\label{unity_root_2}
\eee
$\rP=\pi-\epsilon=\pi-\frac{\pi m}{N}+O\left(\frac{1}{N^2}\right)$ is a
solution of BAE (\ref{BAE_M1}) provided that $0\!<\frac m N\!\ll\! 1$.

Let ${\rP}=-\ir\eta=\ga$, the LHS of (\ref{BAE_M1}) thus becomes
$\eE^{2N\eta-4\eta-2\al_+-2\al_-}$. Obviously, ${\rP}=\ga$ is not a
solution in the general case. However, in the thermodynamic limit
$N\to\infty$, we can always find an integer $m$ to ensure
\bee 
\eta=\frac{\ir\pi m+\al_++\al_-}{N-2}+O\left(\frac1	N\right), \quad N\to\infty,\,\,\,m\to\infty.
\eee 
So ${\rP}=\ga+O\left(\frac1	N\right)$ becomes a solution of $(\ref{BAE_M1})$ in the thermodynamic limit.

\section{$M=1$ current: General case }

{Consider a hermitian Hamiltonian in the easy plane regime
  (\ref{hermiticity_1}) with $\al_{\pm}$, $\theta_\pm$ and $\eta=\ir\gamma$
  being purely imaginary and $\be_+=-\be_-$ being real. The norm of the
  eigenvector $\langle \Psi|$ is
\bee 
\langle \Psi|\Psi\rangle=\left(1+\eE^{-2\be_+}\right)^N\left(\sum_{n\leq m}\left(f_n\,f^\ast_m\,\tilde b_+^{m-n}+f_n^\ast\,f_m\,b_+^{m-n}\right)-\sum_n f_n\,f^\ast_n\right),
\eee
where $f^\ast_n$ is the complex conjugate of $f_n$ and $b_+(x)$ and $\tilde b_+(x)$ are defined in (\ref{def;ab}) and (\ref{def;ab2}) respectively. Then, the current can be obtained by
\begin{align}
{j}^z=\frac{\langle\Psi|\,\mathbf j_l^z|\Psi\rangle}{\langle\Psi|\Psi\rangle}&=\frac{2\sin\gamma}{\ch^2(\be_+)\langle \Psi|\Psi\rangle}\left(-\sum_{n=0}^Nf_{n}f^\ast_{n}+\sum_{\substack{n_1\leq n_2<l\\l<n_1\leq n_2}}\left(f_{n_1}\,f^\ast_{n_2}\,\tilde b_+^{n_2-n_1}+ f^\ast_{n_1}\,f_{n_2}\,b_+^{n_2-n_1}\right)\right.\no\\
&\quad \left.+\sum_{n_1<l<n_2}\left(f_{n_1}\,f^\ast_{n_2}\,\tilde b_+^{n_2-n_1-2}+ f^\ast_{n_1}\,f_{n_2}\,b_+^{n_2-n_1-2}\right)\right).\label{def;current}
\end{align}
The current is independent of the site number.
%When $\ch(\al_-+\be_-+\eta)=0$, i.e., $\beta_-=0,\alpha_-=-\eta\pm\frac{\ir\pi}{2}$, the state $\langle \Psi|$ becomes a spin-helix state and \bee \mathbf{j}=2 \sin\gamma.\eeeSimilarly, when $\beta_+=0,\alpha_+=-\eta\pm\frac{\ir\pi}{2}$, the current also reads\bee \mathbf{j}=2 \sin\gamma.\eee
For a hermitian Hamiltonian, the parameter ${\rP}$ should be real or purely
imaginary. If $\rP$ is real for large $N$, the norm is order $O(N)$ and we
obtain the expression of the current in the leading approximation as
\bee
j^z=\frac{2\sin\gamma}{\ch^2(\be_+)}\left(1-O\left(\frac1N\right)\right).
\eee
If ${\rP}$ is purely imaginary, the norm is order $O(\eE^{2N\abs{\rP}})$ and the current is
\bee j^z=\frac{2\,\sin\gamma}{\ch^2(\be_+)}\left(1-O\left(\eE^{-N\abs{\rP}}\right)\right).
\eee

}

\end{document}